\documentclass[journal=jacsat,manuscript=article]{achemso}

\usepackage[version=3]{mhchem} 
\usepackage{comment}
\usepackage{xcolor}
\usepackage{amsmath}
\usepackage{braket}
\usepackage{amssymb}
\usepackage{relsize}
\usepackage{float}
\usepackage{makecell}
\usepackage{multirow}
\usepackage{setspace}
\usepackage{academicons} 
\usepackage{hyperref}
\hypersetup{   
    colorlinks=true,
    linkcolor=black,
    citecolor=black,
    urlcolor=black
}

\providecommand{\theHpage}{\arabic{page}}
\providecommand{\theHtable}{\arabic{table}}
\providecommand{\theHfigure}{\arabic{figure}}
\providecommand{\theHequation}{\arabic{equation}}
\providecommand{\theHsection}{\arabic{section}}
\providecommand{\theHsubsection}{\arabic{section}.\arabic{subsection}}

\setkeys{acs}{doi=false}

\newcommand{\orcid}[1]{\href{https://orcid.org/#1}{\textcolor{green}{\aiOrcid}}}

\newcommand{\beginsupplement}{%
    \setcounter{table}{0}%
    \renewcommand{\thetable}{S\arabic{table}}%
    \setcounter{page}{1}%
    \renewcommand{\thepage}{S\arabic{page}}%
    \setcounter{figure}{0}%
    \renewcommand{\thefigure}{S\arabic{figure}}%
    \setcounter{equation}{0}%
    \renewcommand{\theequation}{S\arabic{equation}}%
    \setcounter{section}{0}%
    \renewcommand{\thesection}{S\arabic{section}}%
    \setcounter{subsection}{0}%
    \renewcommand{\thesubsection}{S\arabic{section}.\arabic{subsection}}%
    \renewcommand{\theHpage}{SI.\arabic{page}}%
    \renewcommand{\theHtable}{SI.\arabic{table}}%
    \renewcommand{\theHfigure}{SI.\arabic{figure}}%
    \renewcommand{\theHequation}{SI.\arabic{equation}}%
    \renewcommand{\theHsection}{SI.\arabic{section}}%
    \renewcommand{\theHsubsection}{SI.\arabic{section}.\arabic{subsection}}%
}

\author{Marta Monti}
\affiliation[UvA]{Van ’t Hoff Institute for Molecular Sciences, University of Amsterdam, Science Park 904, 1098 XH Amsterdam, The Netherlands}
\email{m.monti@uva.nl}

\author{Alec Noppe}
\affiliation[UvA]{Van ’t Hoff Institute for Molecular Sciences, University of Amsterdam, Science Park 904, 1098 XH Amsterdam, The Netherlands}
\alsoaffiliation[Uva2]{AI4Science Laboratory, University of Amsterdam, Science Park 904, 1098 XH Amsterdam, The Netherlands}

\author{Olivier Lugier}
\affiliation[NanoHybrids]{Nano Hybrids B.V., Science Park 904, 1098 XH Amsterdam, The Netherlands}

\author{Maria B. Brands}
\affiliation[Elexel]{Elexel B.V., Science Park 408, 1098 XH Amsterdam, The Netherlands}

\author{Bernd Ensing}
\affiliation[UvA]{Van ’t Hoff Institute for Molecular Sciences, University of Amsterdam, Science Park 904, 1098 XH Amsterdam, The Netherlands}
\alsoaffiliation[Uva2]{AI4Science Laboratory, University of Amsterdam, Science Park 904, 1098 XH Amsterdam, The Netherlands}
\email{b.ensing@uva.nl}

\title[An \textsf{achemso} demo]
  {Adsorbate-Free Screening of Dissociative Water Adsorption on Nickel Nanoparticles via Graph Neural Networks}

\abbreviations{IR,NMR,UV}

\begin{document}

\let\siaddcontentsline\addcontentsline
\renewcommand{\addcontentsline}[3]{}


\begin{abstract}
Dissociative water adsorption is a key step in heterogeneous and electrochemical catalysis, but on finite Ni nanosystems the OH$^*$ and H$^*$ fragments can occupy many combinations of surface sites, making exhaustive density functional theory (DFT) screening impractical. We introduce a graph neural network (GNN) framework with an adsorbate-free representation: candidate O- and H-binding sites are defined by sets of Ni atoms (\textit{anchors}), and each site pair is encoded as a graph constructed solely from these anchors and their local Ni environment. Using only MACE-relaxed bare Ni geometries as input, separate GNNs predict the DFT-reference adsorption energies of OH$^*$+H$^*$ configurations and whether the proposed binding sites would be retained upon DFT relaxation, eliminating the need to construct or relax adsorbates during screening. Trained across Ni nanoclusters and Ni(111)/Ni(100) surfaces, the energy model achieves a mean absolute error of 0.032 eV on the internal validation set and demonstrates transferability to unseen Ni$_{55}$, Ni$_{91}$, and Ni$_{130}$ clusters and larger surface cells, particularly for relative-energy ranking, although quantitative accuracy varies across systems. These independent tests also support the retention classifier's use as a conservative filter for selecting configurations likely to preserve their proposed binding sites. Application to Ni nanoparticles containing up to $\sim$3000 atoms identifies favorable co-adsorption motifs and shows that lower O-site coordination generally stabilizes the dissociated state, whereas H-site coordination has a weaker influence. The framework thus offers an efficient route to screening adsorption energetics, with potential extensions to other metals and adsorbate combinations.
\end{abstract}

\section{Introduction}
\label{sec:intro}

Water dissociation at metal surfaces is a fundamental step in heterogeneous and electrochemical catalysis, central to processes such as steam methane reforming, the water--gas shift reaction, and the alkaline hydrogen evolution reaction\cite{ke_2020,Lin_2011,subbaraman2011,carrasco2012}. Nickel (Ni) is among the most attractive non-precious metallic catalysts for these reactions\cite{Kim_2008,mahmood2018,Chang_2021}, and its water-dissociation energetics are known from density functional theory (DFT) to depend strongly on surface structure, varying among the low-index facets\cite{mohsenzadeh2014water,Seenivasan_2013}.

\begin{figure}[H]
    \centering
    \includegraphics[width=1.0\linewidth]{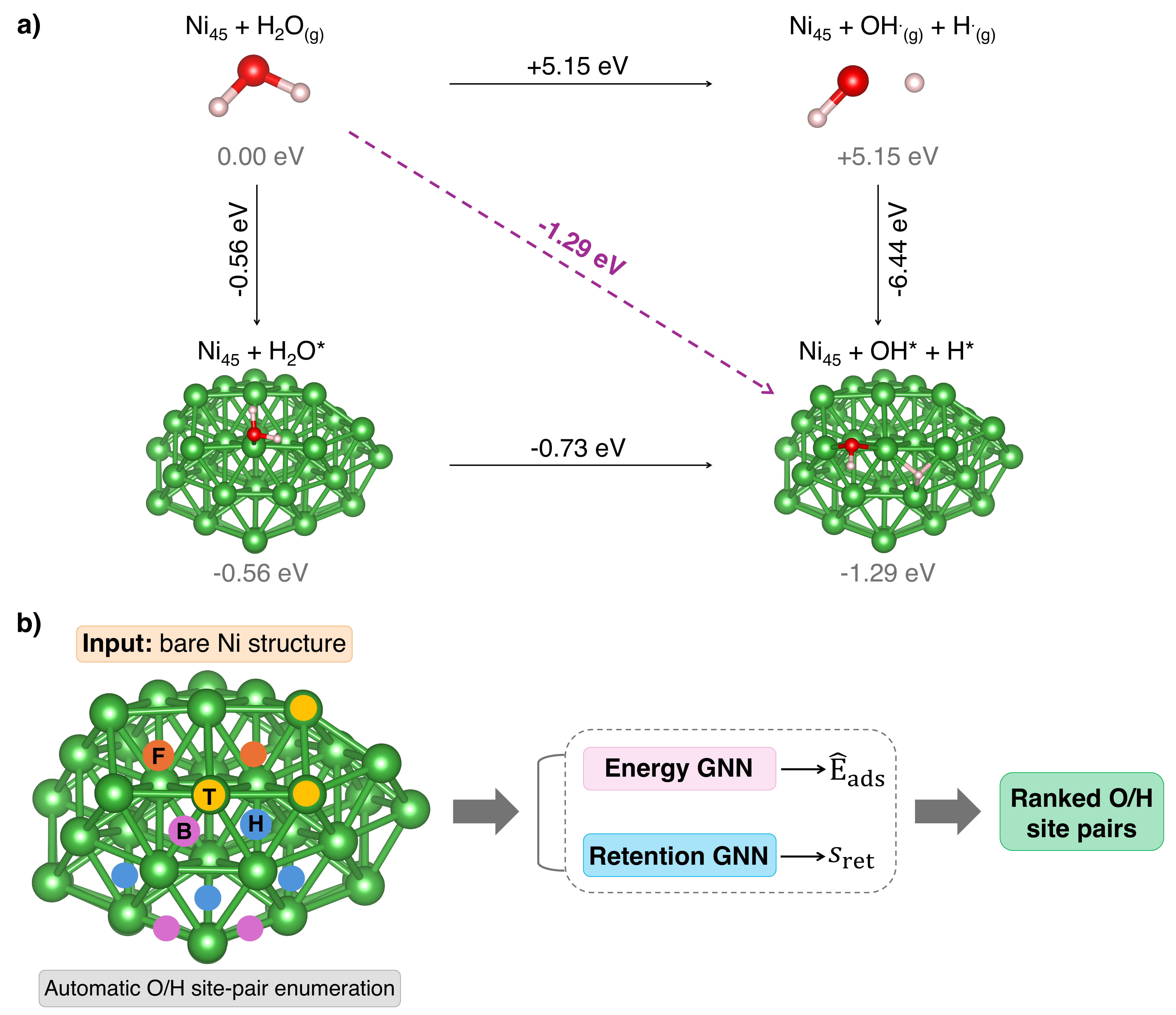}
    \caption{Overview of the anchor-based GNN framework for screening OH$^*$+H$^*$ co-adsorption on bare Ni. (a) Energy diagram illustrating molecular and dissociative water adsorption on a representative Ni$_{45}$ nanocluster. The diagram shows molecular adsorption of H$_2$O followed by dissociation into co-adsorbed OH$^*$ and H$^*$, together with an energy decomposition through gas-phase OH and H radicals and their adsorption onto the Ni surface. All energies are referenced to the bare Ni structure and gas-phase H$_2$O. For clarity, the bare Ni$_{45}$ cluster is not shown in the gas-phase states. The purple dashed arrow highlights the overall dissociative adsorption energy ($\mathrm{E}_{\mathrm{ads}}$) considered in this work (Eq. \ref{eq:eads}). Ni, O, and H atoms are shown in green, red, and white, respectively. (b) Screening workflow. From the bare Ni structure alone, candidate O and H adsorption sites are enumerated automatically and encoded through their Ni anchor atoms, without placing any explicit adsorbate atoms; representative sites of each type---top (T), bridge (B), threefold hollow (H), and fourfold hollow (F)---are labeled, and further markers of the same colors indicate the many additional candidate sites across the surface. Each candidate O/H site pair is passed to two graph neural networks that predict its adsorption energy ($\hat{\mathrm{E}}_{\mathrm{ads}}$) and its motif-retention score ($s_{\mathrm{ret}}$), returning a complete ranking of the enumerated OH$^*$+H$^*$ co-adsorption configurations.}
    \label{fig:intro}
\end{figure}

To quantify the stability of the co-adsorbed OH$^*$ and H$^*$ fragments formed by water dissociation on a catalytic surface (Figure \ref{fig:intro}a), one can measure their adsorption energy ($\mathrm{E}_{\mathrm{ads}}$). Adsorption energetics provide fundamental descriptors of surface reactivity and underpin scaling relationships and volcano-type analyses used to rationalize catalytic trends\cite{Bligaard_2004}. While $\mathrm{E}_{\mathrm{ads}}$ cannot determine reaction rates, it characterizes the stabilization of the dissociation products and is the natural starting point for a structure-resolved description.

Because site- and configuration-specific adsorption energies are difficult to isolate experimentally, the adsorption of OH$^*$, H$^*$, and H$_2$O on Ni has been studied primarily by DFT. Most studies have focused on ideal low-index surfaces, where water binds preferentially atop (T) a single metal atom and the dissociation fragments occupy bridge (B), threefold hollow (H), and fourfold hollow (F) sites\cite{mohsenzadeh2014water,Seenivasan_2013,Zhu_2019,han2025her}. Finite Ni nanosystems (clusters and particles), however, are also of considerable catalytic interest, and their reactivity depends strongly on particle size and morphology\cite{Zhou_2017,Spano_2025}. Unlike ideal flat surfaces, nanoparticles (NPs) expose a broader variety of adsorption environments, including terrace-like regions, under-coordinated edges, and vertices.

Beyond the diversity of individual adsorption sites on Ni catalysts, water dissociation introduces a significant configurational challenge: OH$^*$ and H$^*$ can occupy numerous combinations of nearby binding sites, potentially leading to distinct relaxed co-adsorption configurations. Consequently, the number of configurations to be explored grows rapidly with particle size and structural complexity\cite{Schwalbe-Koda_2025}. Exhaustively evaluating their adsorption energies with DFT therefore becomes computationally prohibitive, making efficient pre-screening strategies particularly valuable.

Machine-learning (ML) models provide a natural route to such pre-screening. Among these approaches, graph neural networks (GNNs) are particularly well suited to atomistic systems, as they learn structure--property relationships directly from atomic environments and enable adsorption-energy predictions orders of magnitude faster than DFT. Building on early crystal- and molecular-graph models\cite{xie2018,schutt2018schnet} and more recent directional and equivariant architectures\cite{gasteiger2021gemnet,liao2024equiformerv2}, large community efforts have established GNNs as powerful surrogates for catalyst screening\cite{chanussot2021oc20,lan2023adsorbml}. ML approaches have also been applied to adsorption-energy prediction on finite metal nanoclusters (NCs) and nanoalloys\cite{jager2020nanoclusters,Otsuka_2026,zhang2026fast}. Graph-based approaches such as ACE-GCN\cite{ghanekar_2022} and the framework developed by Schwalbe-Koda \textit{et al.}\cite{Schwalbe-Koda_2025} further enable extensive sampling of adsorption configurations at high coverages and, in the latter case, under co-adsorption conditions, using explicitly constructed adsorbate--surface structures. Alternative representations have been developed to infer adsorption energetics from clean-surface structures. Malone and Kara predicted site-specific properties of atomic adsorbates directly from periodic-surface geometries, with the atomic species and initial binding site given as inputs.\cite{Malone_2023} Usuga \textit{et al.} predicted adsorption energies on bimetallic alloy surfaces using local adsorbate--surface environments represented through atomic connectivity and physicochemical descriptors derived from the clean surface and gas-phase adsorbate.\cite{usuga_2026} Other approaches include distance-insensitive representations for alloy NCs\cite{Otsuka_2026} and prediction of the global-minimum adsorption energy without specifying individual binding sites\cite{chen2025adsmt}. Although ML-based adsorption-energy prediction has been extensively investigated for extended metal surfaces, analogous applications to finite nanosystems remain comparatively limited. Moreover, existing studies of finite systems have largely considered one adsorbate species at a time\cite{jager2020nanoclusters,zhang2026fast,Otsuka_2026,bang2023machine}, even when accounting for multiple adsorbates at finite coverage, whereas dissociative co-adsorption requires evaluating the combined energetics of distinct fragments across numerous possible binding-site combinations. Systematically screening the full co-adsorption configurational space on an NP surface therefore remains a significant challenge.

To address these challenges, we introduce an anchor-based GNN framework with an adsorbate-free representation for screening dissociative OH$^*$+H$^*$ co-adsorption on finite Ni nanosystems and low-index periodic surfaces. Each adsorption site is defined by a set of Ni atoms, referred to as \textit{anchors}, that characterize its binding geometry (top, bridge, threefold hollow, or fourfold hollow). Motivated by the predominantly local nature of adsorption interactions\cite{prodan2005}, the graph for each candidate O/H site pair is constructed from the anchor atoms themselves together with the neighboring Ni atoms, thereby encoding the local metal environments of both sites without explicitly representing the adsorbates. Starting from the bare Ni structure, the framework automatically enumerates candidate O- and H-binding-site pairs. Separate GNNs then predict, for each site pair, its DFT-reference adsorption energy and whether the pair would be retained under DFT relaxation (Figure \ref{fig:intro}b). Both models use bare Ni geometries relaxed with a machine-learning interatomic potential (MLIP) as their sole structural input and are trained on configurations spanning Ni NCs and the Ni(111) and Ni(100) surface facets, with DFT-reference adsorption energies and motif-retention outcomes as their respective targets. Screening therefore requires neither DFT-relaxed bare structures nor the construction and relaxation of individual adsorbate configurations, enabling application to systems far larger than those practical for exhaustive DFT sampling.

The energy model achieves an MAE of 0.032 eV on the internal validation set and demonstrates transferability to both larger clusters and periodic surface cells, particularly in ranking configurations by relative adsorption energy. The motif-retention classifier further provides a conservative first-pass filter for identifying configurations likely to preserve their initial adsorption environments upon relaxation. Analysis of the predicted adsorption-energy landscapes using the generalized coordination number (GCN)\cite{Calle-Vallejo_2014,callevallejo2014jpcl,Calle-Vallejo_2015} reveals a strong dependence on the coordination of the O-binding site: lower $\mathrm{GCN_O}$ generally stabilizes the co-adsorbed OH$^*$+H$^*$ state, whereas $\mathrm{GCN_H}$ is far less informative. Although demonstrated here for OH$^*$+H$^*$ co-adsorption on Ni, the screening strategy is not inherently restricted to this metal--adsorbate system and could be extended to other metals and adsorbate combinations through appropriate adaptation and training.

\section{Methods}
The main aspects of our proposed workflow are described in this section; more extensive details are provided in the Supporting Information (SI).
\subsection{Dataset generation and model-guided expansion}
\label{sec:dataset_generation}

The bare Ni geometries were taken from the Quantum Cluster Database\cite{manna2023database} and the CSIRO Data Collection\cite{csiro_dataset} and re-optimized at the DFT level (see below); the OH$^*$+H$^*$ co-adsorption configurations built on them, together with their reference adsorption energies, were generated in this work. Candidate configurations were enumerated at T, B, H, and F sites---associated with one, two, three, and four surface Ni atoms, respectively (Figure \ref{fig:intro}b)---and the dataset was grown by a model-guided procedure (see Figure S1): at each stage the current GNN ranked candidates on the next bare NC, and a subset spanning both low- and higher-energy predictions was generated explicitly, relaxed at the DFT level, and added before training the next model. Beginning from M1 (Ni$_{25}$, Ni$_{30}$), this successively introduced Ni$_{35}$, Ni$_{40}$, Ni$_{45}$, and Ni$_{50}$ to give M2--M5 (Figure S1a,b). After relaxation, each configuration's motif was assigned from the Ni atoms coordinating OH$^*$ and H$^*$, which also serve as the anchors through which the GNN represents the adsorbates; full details of site construction, motif assignment, and candidate selection are provided in Sections S1.1--S1.3.

The final M6 model was obtained by fine-tuning M5 on a further-expanded dataset adding two structural isomers of Ni$_{47}$ and periodic slabs exposing the Ni(111) and Ni(100) facets (Figure S1c). These are the two predominant low-energy terminations of Ni ($\gamma_{(111)} < \gamma_{(100)} < \gamma_{(110)}$)\cite{tran2016surface}, make up a large fraction of the exposed area of Ni crystals, and are among the most extensively studied surfaces for Ni-catalyzed water dissociation and hydrogen evolution\cite{mohsenzadeh2014water,Zhu_2019,han2025her}. The more open Ni(110) and stepped surfaces, which are structurally more complex and more costly to sample, are not considered here and represent a natural extension of the model. Surface candidates were generated on the exposed upper slab layer, with all geometrical operations evaluated under periodic boundary conditions and the same M5-guided selection.

\subsection{Electronic-structure calculations}
\label{sec:dft_calc}

The bare Ni structures were optimized with spin-polarized DFT in Quantum ESPRESSO\cite{QE_r1,QE_r2} using the PBE exchange--correlation functional\cite{PBE}, an ultrasoft pseudopotential for Ni\cite{vanderbilt1990soft} and optimized norm-conserving Vanderbilt pseudopotentials for O and H\cite{ONCV}, with Marzari--Vanderbilt smearing\cite{MV_smear} (0.01 Ry) and plane-wave cutoffs of 45 and 360 Ry for the wavefunctions and charge density, respectively. The NCs were placed in cubic supercells (20--25 \AA, depending on NC size) with the Brillouin zone sampled at the $\Gamma$ point, and the Ni+OH$^*$+H$^*$ configurations were relaxed with identical settings. Geometries were optimized with the Broyden--Fletcher--Goldfarb--Shanno (BFGS) algorithm\cite{Broyden1970,Fletcher1970,Goldfarb1970,Shanno1970} until the change in total energy and the maximum residual force fell below $1.0\times10^{-4}$ Ry and $5.0\times10^{-4}$ Ry/Bohr, respectively. The periodic Ni(111) and Ni(100) surfaces were modeled as four-layer slabs---using an orthorhombic surface cell for Ni(111)---with the cell length along the surface normal set so that the separation between periodic slab images exceeded $\sim$18 \AA, and the Brillouin zone sampled with a $\mathrm{k_x}\times\mathrm{k_y}\times1$ Monkhorst--Pack\cite{Monkhorst_1976} grid chosen according to the lateral cell dimensions. The clean slabs were fully relaxed, whereas for the Ni+OH$^*$+H$^*$ configurations the bottom two Ni layers were held fixed at their optimized positions and the remaining atoms were allowed to relax.

Total energies entering the electronic adsorption energy were obtained from single-point calculations in which the kinetic-energy cutoffs were increased to 60 and 480 Ry and Grimme's D3 dispersion correction with Becke--Johnson damping\cite{Grimme_2010,Grimme_2011} was added, applied consistently to all three terms of
\begin{equation}
    \mathrm{E}_{\mathrm{ads}} =
    \mathrm{E}_{\mathrm{Ni+ads}}
    -\mathrm{E}_{\mathrm{Ni}}
    -\mathrm{E}_{\mathrm{ref}},
    \label{eq:eads}
\end{equation}
where $\mathrm{E}_{\mathrm{Ni+ads}}$, $\mathrm{E}_{\mathrm{Ni}}$, and $\mathrm{E}_{\mathrm{ref}}$ are the total energies of the Ni system with the adsorbed species, the corresponding bare Ni system, and the gas-phase reference molecule, respectively. For the dissociated OH$^*$+H$^*$ state considered here, the reference is a single geometry-optimized gas-phase H$_2$O molecule, so that $\mathrm{E}_{\mathrm{ads}}$ measures the energetics of the overall dissociative water-adsorption process. Zero-point-energy (ZPE) corrections were not included: their evaluation would require a vibrational analysis for each of the $\sim$2000 configurations in the dataset, which is computationally expensive. Previous studies of water dissociation on low-index Ni surfaces have reported substantial ZPE corrections to OH$^*$+H$^*$ adsorption energies, on the order of 0.3--0.4 eV\cite{mohsenzadeh2014water}. However, because ZPE corrections depend on the chosen reference state, values computed against isolated OH and H cannot be transferred directly to our H$_2$O-referenced definition. When comparing different adsorption configurations on the same Ni structure, the reference contributions cancel, and only differences in the vibrational corrections of the adsorbed states affect their relative stability. Although these differences are expected to be smaller than the absolute corrections, their magnitude has not been evaluated and could influence the ordering of configurations with similar energies. We therefore treat this as a known limitation of the absolute $\mathrm{E}_{\mathrm{ads}}$ values.

\subsection{Graph neural network for adsorption-energy prediction}

The adsorption-energy model predicts the stability of a candidate OH$^*$+H$^*$ configuration directly from the corresponding bare Ni structure and an encoding of the candidate adsorption sites,
\begin{equation}
    \hat{\mathrm{E}}_{\mathrm{ads}} =
    f_{\theta}\left(
    \mathcal{G}_{\mathrm{Ni_{bare}}},
    \mathcal{A}_{\mathrm{O/H}}
    \right),
    \label{eq:gnn_mapping}
\end{equation}
where $\mathcal{G}_{\mathrm{Ni_{bare}}}$ is a (local) graph built from the bare Ni geometry, $\mathcal{A}_{\mathrm{O/H}}$ encodes the Ni atoms defining the candidate O and H sites (the anchor atoms), and $f_{\theta}$ is the learned GNN. Explicit adsorbate coordinates are therefore not required at inference: during training the relaxed geometry serves only to identify the O and H anchors, from which a local subgraph of the surrounding bare Ni is constructed, so the configuration enters solely through the anchor encoding. The network applies edge-conditioned message passing\cite{gilmer2017mpnn} with a learned edge-gating mechanism\cite{satorras2021egnn}, followed by a mixed global--local (anchor-aware) pooling that combines a global representation of the local Ni environment with separate O- and H-centered representations, and a fully connected head mapping the pooled representation to $\hat{\mathrm{E}}_{\mathrm{ads}}$. It was implemented in PyTorch\cite{paszke2019pytorch} and PyTorch Geometric\cite{fey2019pyg}; graph construction, node features, and the full architecture are detailed in Sections S1.4--S1.5.

Models M1--M5 were trained on DFT-relaxed bare geometries by minimizing the mean-squared error to the DFT adsorption energies (Section S1.6), and M6 was obtained by fine-tuning M5 on the expanded nanocluster-plus-surface dataset with the architecture held fixed (Section S1.7). To remove the need for a DFT relaxation of the bare catalyst when applying the model to larger systems, the full M5--M6 workflow was repeated using bare geometries relaxed with the MACE-MH-1 MLIP (\texttt{omat\_pbe} head)\cite{batatia2022mace,batatia2025crosslearning}, started from the same unrelaxed geometries and optimized with FIRE\cite{bitzek2006structural} to a maximum force of $0.02$ eV\,\AA$^{-1}$ at fixed cell. We refer to the two resulting versions as M6-DFT and M6-MACE, respectively. The adsorption-energy targets remained the DFT reference values in both cases, and M6-MACE---which requires no bare-catalyst DFT relaxation---is used for all subsequent screening and structural analyses. The architecture and training choices were fixed through a series of controlled tests, as discussed in detail in Section S2 (Figures S2--S5, and Table S1).

\subsection{Graph neural network for motif-retention prediction}

Alongside the energy model, we trained a binary GNN classifier to predict whether a candidate OH$^*$+H$^*$ configuration retains its initial adsorption site upon DFT relaxation. It uses the same type of encoder and the same anchor-based, adsorbate-free graph input, with separate O- and H-centered pooling and independent output heads for nanocluster and periodic-surface configurations; a configuration is labeled retained only when both the motif and the complete O- and H-anchor sets are unchanged after relaxation. The classifier was trained with a binary cross-entropy loss, and the sigmoid output serves as a retention score. As for the energy model, it was established on DFT-relaxed inputs and then retrained on the corresponding MACE-relaxed geometries, with the DFT-defined labels and the train/validation split held fixed; full details are given in Section S3 (Tables S2--S5).

\subsection{Evaluation metrics}

Predictive performance is reported as the mean absolute error (MAE), root-mean-square error (RMSE), and coefficient of determination (R$^2$). Because the model is intended to rank competing adsorption configurations on a given catalyst, we additionally report the relative adsorption energy $\Delta \mathrm{E}_{\mathrm{ads}}$, which measures how accurately differences in stability are preserved. Within each system, energies were referenced to the lowest-energy DFT configuration, $i_{\mathrm{DFT,min}}$:
\begin{align}
\Delta \mathrm{E}_{\mathrm{ads}}^{\mathrm{DFT}}(i)
&=
\mathrm{E}_{\mathrm{ads}}^{\mathrm{DFT}}(i)
-
\mathrm{E}_{\mathrm{ads}}^{\mathrm{DFT}}(i_{\mathrm{DFT,min}}),
\label{eq:relative_dft}
\\
\Delta \mathrm{E}_{\mathrm{ads}}^{\mathrm{pred}}(i)
&=
\mathrm{E}_{\mathrm{ads}}^{\mathrm{pred}}(i)
-
\mathrm{E}_{\mathrm{ads}}^{\mathrm{pred}}(i_{\mathrm{DFT,min}}),
\label{eq:relative_pred}
\end{align}
where the predicted relative energies are referenced to the prediction for the same configuration that defines the DFT minimum. This removes uniform offsets between predicted and reference adsorption energies while preserving the predicted energy differences relative to the DFT-reference minimum.

\section{Results and Discussion}
\subsection{Model accuracy using MACE-relaxed bare structures}

\begin{figure}[H]
    \centering
    \includegraphics[width=0.9\linewidth]{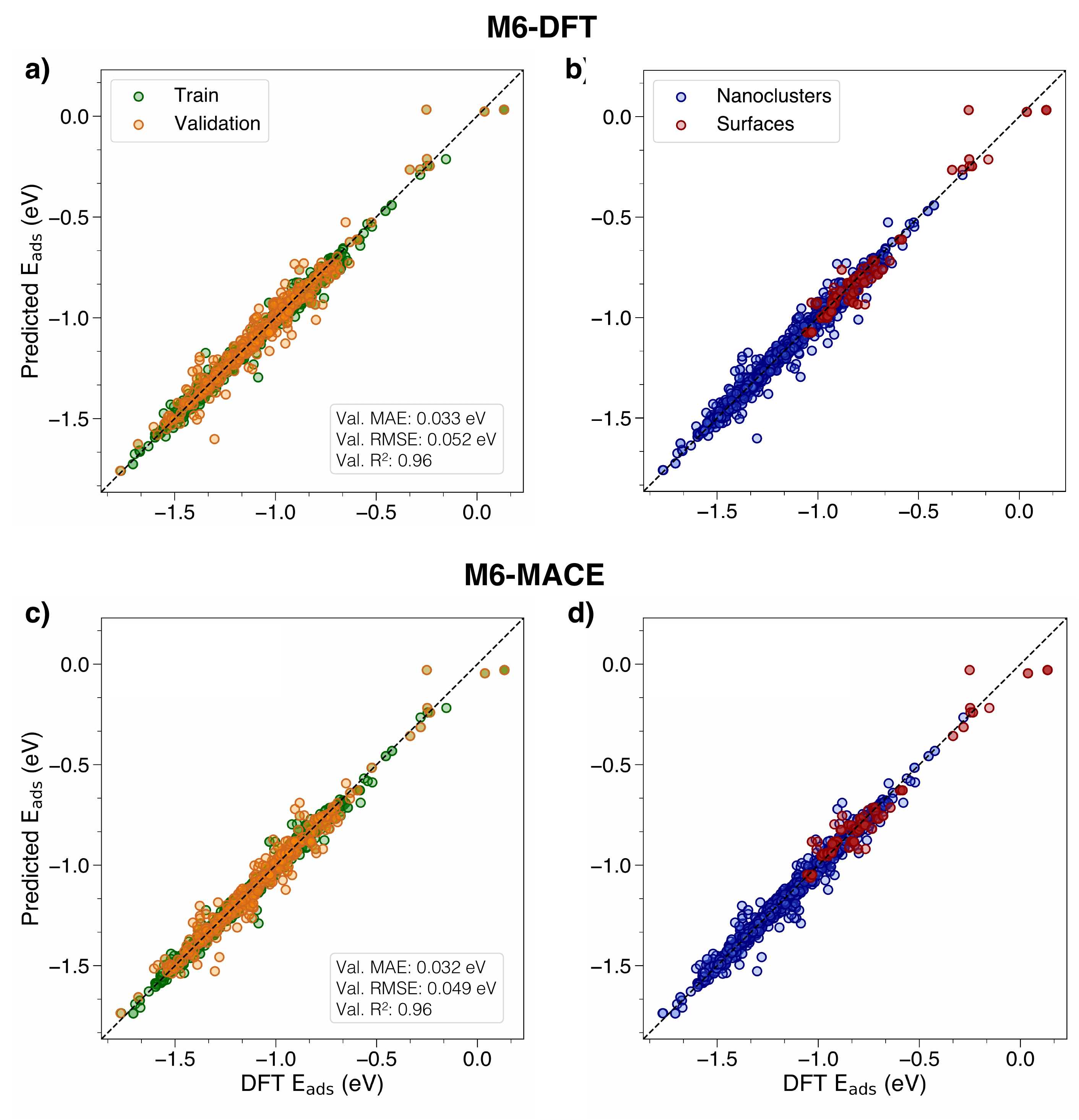}
    \caption{Comparison of the in-distribution performance of the M6-DFT and M6-MACE adsorption-energy models. Predicted versus DFT-reference adsorption energies, $\mathrm{E}_{\mathrm{ads}}$, are shown using (a,b) DFT-relaxed and (c,d) MACE-relaxed bare Ni geometries as graph inputs. Panels (a,c) distinguish training (green) and validation (orange) configurations, while panels (b,d) distinguish nanocluster (blue) and periodic-surface (red) configurations. The reported mean absolute error (MAE), root-mean-square error (RMSE), and coefficient of determination (R$^2$) refer to the validation set; the dashed diagonal line indicates perfect agreement.}
    \label{fig:m6_parity}
\end{figure}

The final M6 model was obtained by fine-tuning the NC-trained model on an expanded dataset that adds two Ni$_{47}$ structural isomers and the periodic Ni(111) and Ni(100) surfaces (Figure S1), extending the range of structural environments represented during training. The resulting dataset spans a broad range of adsorption energetics, with DFT $\mathrm{E}_{\mathrm{ads}}$ from approximately $-1.80$ to $-0.40$ eV for the NCs and mostly between $-1.00$ and $-0.50$ eV for the periodic surfaces, with a small fraction of the latter extending to less negative values. Using DFT-relaxed bare structures as inputs (M6-DFT), the model predicts OH$^*$+H$^*$ adsorption energies on internal validation configurations with an MAE of 0.033 eV, an RMSE of 0.052 eV, and $\mathrm{R}^2 = 0.96$ (Figure \ref{fig:m6_parity}a). Comparable accuracy is observed for both nanocluster and periodic-surface configurations (Figure \ref{fig:m6_parity}b). As the validation configurations are drawn from systems represented during training, and this set was also used for early stopping and model selection, these results characterize predictive accuracy within the sampled structural domain, while transferability to previously unseen systems is assessed independently in the following subsection.

A central feature of the representation is that the input graph is built solely from the bare metal geometry and the adsorption-site anchors, without explicit adsorbate coordinates. Consequently, screening requires only an optimized bare metal structure, rather than the construction and relaxation of individual adsorbate configurations. We therefore investigate whether the bare Ni geometries can be obtained using an MLIP instead of DFT, enabling the application of the framework to larger, more realistic NPs for which DFT geometry optimization would be computationally prohibitive. Repeating the complete M5-to-M6 training workflow using bare geometries relaxed with the MACE-MH-1 potential\cite{batatia2022mace,batatia2025crosslearning} (M6-MACE) reproduces the validation performance of the DFT-geometry model (M6-DFT), with a validation MAE of 0.032 eV and an RMSE of 0.049 eV, compared to 0.033 eV and 0.052 eV, and an identical $\mathrm{R}^2$ of 0.96 (Figure \ref{fig:m6_parity}c). The two variants share the same architecture, hyperparameters, and data splits, allowing the influence of the bare-geometry source on model performance to be assessed directly. The agreement holds separately for NCs and surfaces (Figure \ref{fig:m6_parity}d), demonstrating that MACE-relaxed geometries retain the structural information needed to reproduce DFT-reference adsorption energies with comparable accuracy. A system-resolved analysis of M6-MACE (Figures S6--S7 and Table S6) further confirms that the validation MAEs for absolute adsorption energies remain below 0.09 eV for all individual NCs and surfaces. 

To benchmark the predictive performance of our model against conventional adsorption-site approaches, we compared M6-MACE with Ridge and XGBoost regression models\cite{Hoerl_1970,Chen_2016} using the co-adsorption motif, the O- and H-site GCNs, and the site--site distance (details in Section S5). On the internal validation set, M6-MACE achieves an MAE of 0.032 eV, compared with 0.111 eV for Ridge and 0.072 eV for XGBoost (Figure S8). The improvement is particularly pronounced for NCs, indicating that the complete GNN framework captures structural information not fully represented by these conventional descriptors. The performance of these models on the external test systems is provided in Figure S9 and discussed in the following subsection.

The motif-retention classifier likewise preserved its discrimination performance when retrained using MACE-relaxed rather than DFT-relaxed bare Ni geometries as inputs. On its internal validation set, the MACE-input classifier achieved an average precision (AP) of 0.911 and a ROC--AUC of 0.972 (Table S5). At the screening threshold of $s_{\mathrm{ret}}=0.80$, chosen to favor precision over recall (Section S3.4), it yielded a precision of 0.933, a recall of 0.680, and a specificity of 0.990. The classifier therefore acts as a conservative first-pass filter, accepting fewer valid configurations in exchange for limiting the number of migrating candidates sent to DFT relaxation.

We therefore adopt the final M6-MACE energy model together with the MACE-input motif-retention classifier for all subsequent analyses. Their reliance on MACE-relaxed bare structures enables the screening of NPs well beyond the size range for which exhaustive DFT evaluation of adsorption configurations is computationally practical, as explored in the following transferability and structural analyses.

\subsection{Transferability to larger nanoclusters and surfaces}

\begin{figure}[H]
    \centering
    \includegraphics[width=1.0\linewidth]{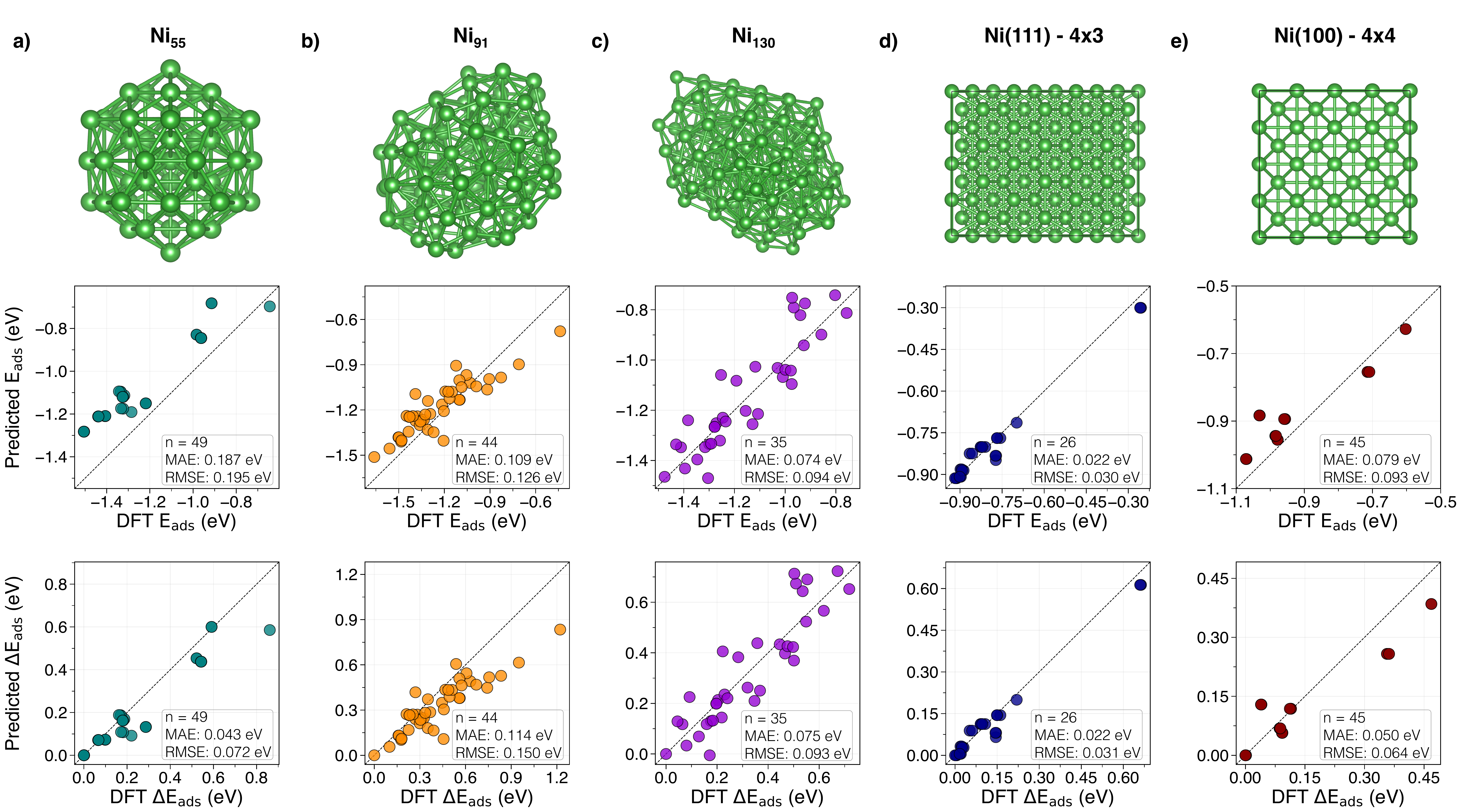}
    \caption{Comparison between DFT-reference and GNN-predicted adsorption energies for selected configurations of three external test nanoclusters: (a) Ni$_{55}$, (b) Ni$_{91}$, and (c) Ni$_{130}$; and two extended periodic surfaces: (d) Ni(111)-4$\times$3 and (e) Ni(100)-4$\times$4. The top row shows the corresponding bare Ni structures, while the middle and bottom rows compare absolute ($\mathrm{E}_{\mathrm{ads}}$) and relative ($\Delta\mathrm{E}_{\mathrm{ads}}$) adsorption energies, respectively. Relative energies are referenced to the lowest-energy DFT configuration within the selected subset for each system, according to Eqs. \ref{eq:relative_dft}--\ref{eq:relative_pred}. Each panel reports the number of configurations, mean absolute error (MAE), and root-mean-square error (RMSE); the dashed diagonal lines indicate perfect agreement.}
    \label{fig:transferability}
\end{figure}

To assess generalization beyond the validation set, we evaluated both models on an external test set of three larger finite Ni systems (Ni$_{55}$, Ni$_{91}$, and Ni$_{130}$) and two larger periodic surfaces, Ni(111)-4$\times$3 and Ni(100)-4$\times$4 (Figure \ref{fig:transferability}, top row). None of these systems was used for training, hyperparameter tuning, or model selection, so they provide an independent assessment of transferability to unseen structures. For each system, the framework ranks candidate OH$^*$+H$^*$ configurations directly from the bare Ni structure, assigning each a predicted adsorption energy and motif-retention score from its anchoring environment, without explicitly constructing the adsorbates. A subset spanning the predicted energy range was then explicitly built and relaxed at the DFT level, during which the adsorbates may migrate to different sites. This provides two complementary assessments: the DFT adsorption energies are compared with the model predictions for the final, relaxed motifs, and the observed retention or migration is compared with the classifier. We evaluate both absolute ($\mathrm{E}_{\mathrm{ads}}$) and relative ($\Delta\mathrm{E}_{\mathrm{ads}}$) adsorption energies, with the latter referenced to the lowest-energy DFT configuration of each subset (Figure \ref{fig:transferability}, middle and bottom rows, respectively). The corresponding ranking correlations are reported in Figure S10; the retention classifier is discussed separately below.

Performance varies across the three finite systems (Figure \ref{fig:transferability}a--c). For Ni$_{55}$, M6-MACE reproduces the relative energies with an MAE of 0.043 eV (49 configurations)--far below the absolute-energy MAE of 0.187 eV--showing that the model captures energy differences between configurations despite a near-uniform offset in their absolute values. For Ni$_{91}$, the relative-energy MAE rises to 0.114 eV (44 configurations), with the largest deviations occurring for the least favorable states, where predicted energies saturate near 0.6 eV while the corresponding DFT values extend up to $\sim$0.9 eV. The least favorable configuration of the set ($\Delta\mathrm{E}_{\mathrm{ads}}\sim1.2$ eV) is likewise underestimated, with a predicted value of $\sim$0.83 eV. Low- and high-energy ranges nonetheless remain well separated, and the absolute energies are reproduced with an MAE of 0.109 eV. For Ni$_{130}$, the relative-energy MAE is 0.075 eV (35 configurations), with comparable absolute accuracy (MAE = 0.074 eV). Each subset included configurations among the lowest-energy candidates predicted to retain their motifs, and DFT confirmed both their favorable adsorption energies and motif retention, demonstrating the model's ability to identify promising adsorption sites. In all three NCs, the lowest-energy configurations are BH motifs. The corresponding Spearman rank correlations of 0.95, 0.87, and 0.92 (Figure S10a--c) further confirm that the energetic ordering is largely preserved even where quantitative accuracy decreases.

The periodic surfaces show similarly encouraging results. M6-MACE describes Ni(111)-4$\times$3 very accurately, with relative and absolute MAEs of 0.022 eV over 26 configurations (Figure \ref{fig:transferability}d) and a Spearman $\rho = 0.90$ (Figure S10d). The smaller number of configurations in this subset reflects the cost of this 96-atom cell, which---unlike the nanoclusters---requires sampling over a $k$-point grid, together with the high symmetry of the flat facet, under which many candidate site pairs are equivalent. The test set also contains two distorted HT configurations which, as expected, are less stable on the (111) facet ($\Delta\mathrm{E}_\mathrm{ads}>0.6$ eV).\cite{mohsenzadeh2014water} For Ni(100)-4$\times$4, we report a relative MAE of 0.050 eV and an absolute MAE of 0.079 eV over 45 configurations (Figure \ref{fig:transferability}e). As on Ni(111), the high symmetry of this flat facet makes many of these configurations equivalent, so they coincide as overlapping points. Here, the deviations are concentrated among energetically close configurations: BF configurations lying $\sim$0.04 eV above the DFT minimum are predicted to be less favorable than FF configurations located $\sim$0.09 eV above the minimum, reversing their DFT ordering and lowering the rank correlation ($\rho = 0.31$, Figure S10e). The higher-energy configurations ($\Delta\mathrm{E}_\mathrm{ads}\geq 0.3$ eV), in contrast, sit close to the parity line and correspond to BB motifs. Notably, the most stable retained configurations are correctly identified on both facets, with the predicted minima---HH on Ni(111) and FF on Ni(100)---matching the site preferences established by previous DFT studies of OH and H adsorption on these surfaces\cite{mohsenzadeh2014water,han2025her}. Furthermore, the minimum $\mathrm{E}_{\mathrm{ads}}$ is more negative on Ni(100) than on Ni(111), consistent with the stronger stabilization of water-dissociation products on the former facet reported by Mohsenzadeh \textit{et al.}\cite{mohsenzadeh2014water}. Overall, these results demonstrate the transferability of the learned adsorption energetics from smaller training cells to larger periodic systems, correctly reproducing the preferred adsorption motifs on both facets.

The comparison with Ridge and XGBoost further confirms the advantages of M6-MACE on the external test systems (Figure S9). The GNN achieves the lowest absolute- and relative-energy MAEs for the three NCs and Ni(111)-4$\times$3, whereas on Ni(100)-4$\times$4 the descriptor-based models perform comparably or slightly better (Ridge: 0.055 and 0.045 eV versus 0.079 and 0.050 eV for M6-MACE). These results highlight the ability of the present framework to resolve adsorption-energy differences across structurally diverse environments, particularly for NCs.

\begin{figure}[H]
    \centering
    \includegraphics[width=0.8\linewidth]{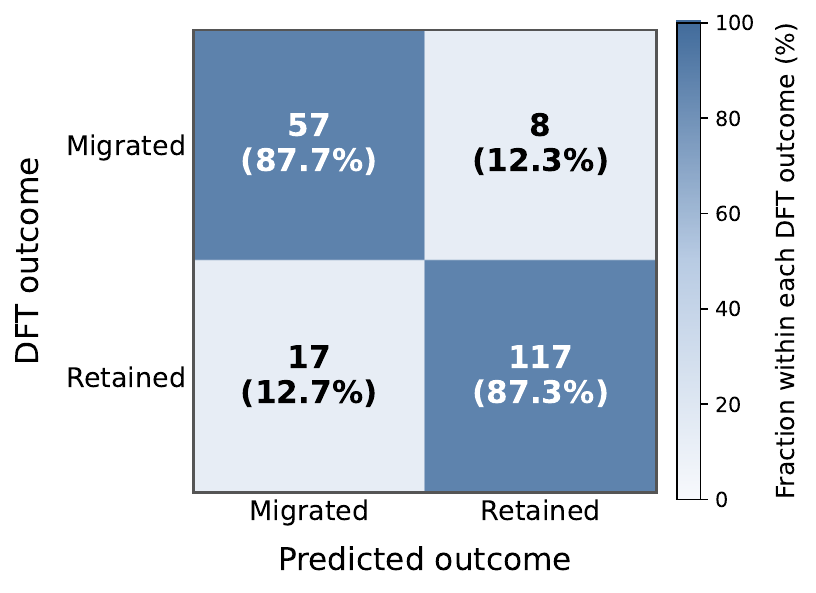}
    \caption{Confusion matrix for motif-retention predictions on the external test systems ($N=199$). The DFT outcome is classified as retained when the initial adsorption motif and the complete O- and H-anchor sets are preserved following DFT relaxation. Predicted outcomes were assigned using a retention-score threshold of $s_{\mathrm{ret}}=0.80$. Values report the number of configurations and, in parentheses, the percentage within each DFT outcome; the diagonal percentages therefore represent class-specific recall. }
    \label{fig:retention}
\end{figure}

The DFT calculations also provide an independent evaluation of the motif-retention classifier across the external systems. Identifying which proposed adsorbate configurations survive relaxation is itself a recognized challenge in high-throughput screening, as initial placements frequently relax to different sites or motifs, enlarging the configuration space that must be explored\cite{gu_2022,lan2023adsorbml}. Aggregated across all five external systems (Figure \ref{fig:retention}), the classifier recovers 117 of the 134 motif-retaining configurations and correctly flags 57 of the 65 migrating ones (87.3\% and 87.7\% recall, respectively), while 117 of the 125 cases it predicts to be retained are confirmed by DFT (93.6\% precision). The retained-class precision is close to that obtained on the internal validation set (93.3\%), indicating that the classifier transfers to these unseen configurations without significant loss of performance. The system-resolved results (Table S7) reveal high retained-class precision for Ni$_{55}$, Ni$_{91}$, and both periodic surfaces (91.7--97.5\%), while performance deteriorates for Ni$_{130}$, which shows both the lowest precision (82.6\%) and the lowest classification accuracy (77.1\%). Across all systems, the errors are dominated by retained configurations classified as migrating (17 cases) rather than by migrating configurations accepted as retained (8 cases): the classifier overlooks some stable adsorption sites but rarely accepts migrating ones. These results support its use as a conservative first-pass filter for identifying adsorption sites likely to preserve their initial binding motifs.

Overall, M6-MACE can explore dissociative water adsorption on previously unseen Ni systems without requiring DFT-relaxed bare structures. Owing to the local nature of the representation, transferability is expected to depend primarily on the similarity of adsorption-site environments to those represented during training, rather than on the overall system size. Across the independently DFT-evaluated test configurations, the model generally reproduces relative energetic trends, discriminating between strong and weak adsorption regimes, although accuracy varies with surface morphology. We note that these errors are evaluated on subsets spanning the predicted energy range and motif diversity of each system, rather than the full configuration space. Consequently, they characterize the predictive performance on the evaluated configurations rather than providing an unbiased estimate of the error across all possible adsorption sites. On these external systems, the absolute-energy MAEs (0.022--0.187 eV) are of the same order of magnitude as those reported for other ML approaches to adsorption-energy prediction on nanoclusters and nanoalloys\cite{Otsuka_2026,bang2023machine,jager2020nanoclusters,zhang2026fast}. Particularly relevant is the recent work of Otsuka \textit{et al.}\cite{Otsuka_2026}, who reported MAEs of 0.105--0.122 eV when transferring their models to larger, previously unseen 85-atom alloy clusters. Although differences in the adsorption systems and prediction targets preclude a direct quantitative comparison, these results provide a useful reference for assessing the quality of our results. Combined with the retention classifier, the model thus offers a computationally efficient route to screen adsorption-energy landscapes across structurally diverse finite Ni nanosystems and periodic surfaces.

Beyond assessing predictive performance on these systems, we evaluated the practical applicability of the framework to systems for which exhaustive DFT calculations of co-adsorption configurations would be computationally prohibitive. We therefore applied the screening workflow to larger MACE-relaxed NPs containing between 218 and 3007 Ni atoms. Predictions for these larger systems are not DFT-validated and are interpreted as model inferences. Candidate O/H site pairs are enumerated and ranked directly from the bare geometries, without requiring DFT calculations during screening. To characterize the computational cost of the current implementation, we measured wall-clock times for representative systems ranging from Ni$_{55}$ to Ni$_{3007}$ (Table S8). Screening required approximately 5 min for Ni$_{55}$ (19,680 configurations) and 255 min for Ni$_{3007}$ (146,914 configurations). The number of screened configurations does not increase monotonically with particle size: Ni$_{1141}$ yields more candidate pairs (181,876) than Ni$_{3007}$, reflecting differences in surface morphology and in the spatial distribution of adsorption-site motifs. The reported timings include candidate-site enumeration, filtering, pre-processing, GNN inference, and subsequent ranking; the preceding MACE relaxation of the bare structures, which took on the order of minutes, is not included. They therefore characterize the complete current workflow and do not scale solely with the number of configurations evaluated by the GNNs.

\subsection{Structural determinants of adsorption-site stability}

We next examine how predicted adsorption stability varies with local structure. We focus this analysis on finite Ni nanosystems, whose exposed sites span a broad range of local coordination environments, whereas the high symmetry of the ideal Ni(111) and Ni(100) terminations yields a more limited range. For each finite Ni system, the analysis includes the complete set of enumerated candidate configurations predicted to retain their anchoring sites. We first examine the Ni$_{55}$, Ni$_{91}$, and Ni$_{130}$ NCs, for which model performance was independently benchmarked against DFT using selected configurations, and then apply the same analysis to the larger, prediction-only NPs.

\begin{figure}[H]
    \centering
    \includegraphics[width=1.0\linewidth]{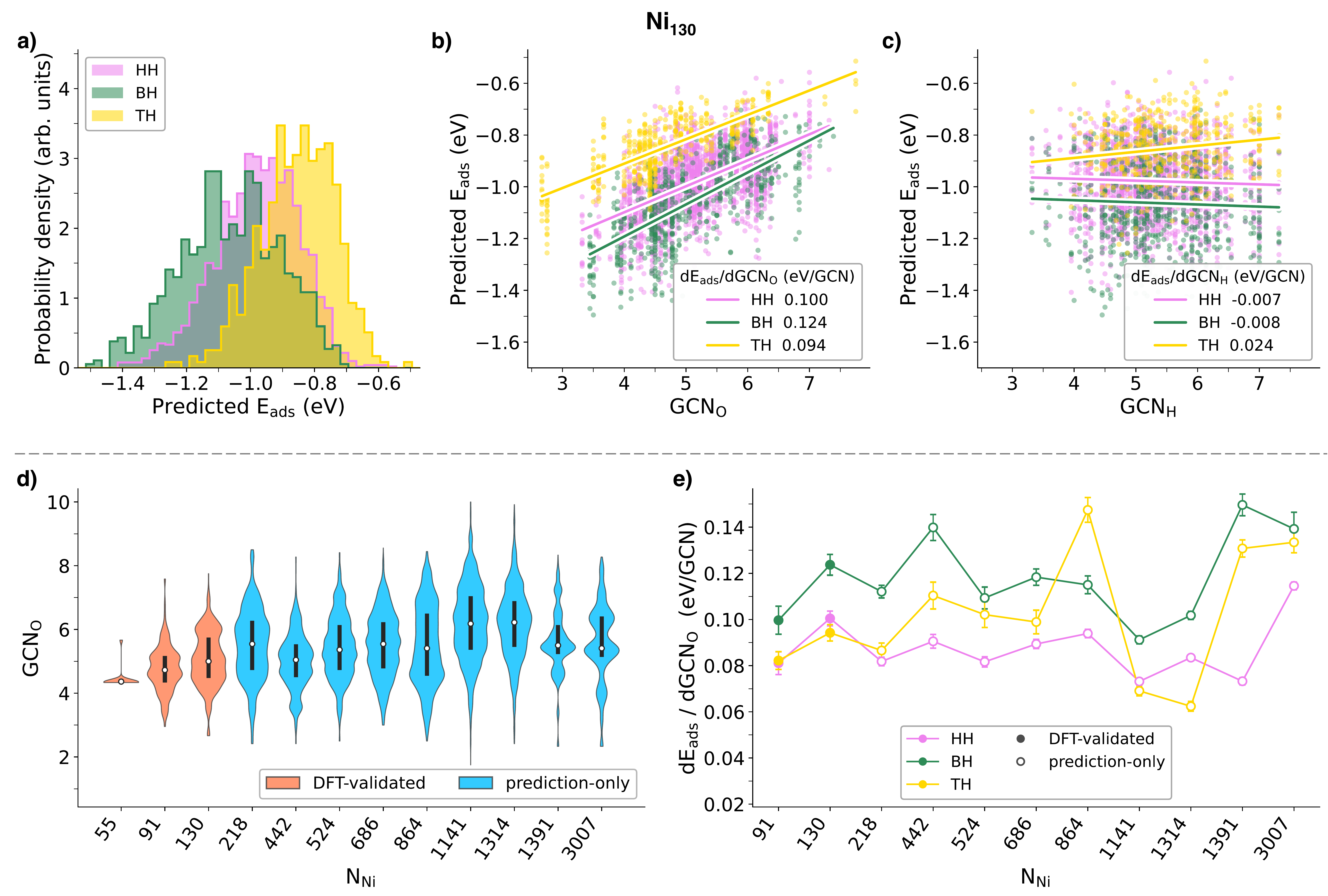}
    \caption{Structural analysis of OH$^*$+H$^*$ adsorption energies predicted by M6-MACE for configurations passing the retention filter. (a) Distribution of predicted adsorption energies, $\mathrm{E}_{\mathrm{ads}}$, for Ni$_{130}$, resolved by co-adsorption motif (BH, HH, TH) and normalized per motif. (b,c) Predicted $\mathrm{E}_{\mathrm{ads}}$ for Ni$_{130}$ as a function of the generalized coordination number of (b) the O site ($\mathrm{GCN_O}$) and (c) the H site ($\mathrm{GCN_H}$), colored by motif; lines are per-motif linear fits, with the corresponding slopes given in the legend. (d) $\mathrm{GCN_O}$ distribution among configurations predicted to retain their adsorption motifs as a function of Ni size (Ni$_{55}$--Ni$_{3007}$); orange denotes the DFT-validated nanoclusters and blue the prediction-only nanoparticles. (e) Motif-resolved slopes $\mathrm{d}\mathrm{E}_{\mathrm{ads}}/\mathrm{d}\mathrm{GCN_O}$ across Ni sizes (filled: DFT-validated; open: prediction-only); error bars indicate the standard errors of the fitted slopes.}
    \label{fig:structural}
\end{figure}

For these DFT-validated systems, the retained configurations are dominated by three co-adsorption motifs---HH, BH, and TH---while other motif families are present but statistically minor (typically $<5\%$). Their predicted adsorption-energy distributions overlap substantially, particularly for BH and HH, with BH reaching the most negative values. In contrast, the TH distribution is shifted toward less negative adsorption energies, although considerable overlap with the other motifs remains (Figure \ref{fig:structural}a). This behavior is illustrated here for Ni$_{130}$ and is reproduced on Ni$_{91}$ (not shown); Ni$_{55}$ is instead a special case, its high-symmetry magic-number structure giving a narrow, nearly degenerate set of environments.

To identify the structural factors underlying these energetic trends, we related $\mathrm{E}_{\mathrm{ads}}$ to the GCN of the O- and H-binding sites\cite{Calle-Vallejo_2014,callevallejo2014jpcl,Calle-Vallejo_2015}. The predicted adsorption energy increases systematically with the coordination of the O site ($\mathrm{GCN_O}$) for every motif, whereas the dependence on the coordination of the H site ($\mathrm{GCN_H}$) is considerably weaker, with fitted slopes close to zero (Figure \ref{fig:structural}b,c; Ni$_{55}$ and Ni$_{91}$ in Figure S11). Since more negative adsorption energies indicate greater stabilization, this positive correlation reveals that less coordinated O-binding environments, associated with under-coordinated Ni atoms, favor stronger binding of the dissociated OH$^*$+H$^*$ state. 

Because $\mathrm{GCN_O}$ and $\mathrm{GCN_H}$ are not fully independent---and because co-adsorption could introduce a dependence on both fragments and their lateral arrangement---we further investigated their respective contributions through a multivariable regression and a drop-one-predictor analysis (Figure S12). The multivariable fit confirms the dominant contribution of $\mathrm{GCN_O}$, whose coefficient is substantially larger than those associated with $\mathrm{GCN_H}$ and their interaction. Consistently, the drop-one-predictor analysis attributes the largest unique contributions to $\mathrm{GCN_O}$ and adsorption motif, whereas $\mathrm{GCN_H}$ and site--site distance provide negligible additional explanatory power. These results indicate that variations in the predicted stability of co-adsorbed OH$^*$+H$^*$ are predominantly associated with the coordination environment of the hydroxyl-binding site, whereas the H-site coordination provides comparatively little additional information.

Previous studies have established relationships between GCNs and adsorption energies for individual adsorbates, including oxygen- and hydrogen-containing species on metal NPs and extended surfaces\cite{Calle-Vallejo_2014,callevallejo2014jpcl,Calle-Vallejo_2015,Calle-Vallejo_2023}. In the context of water dissociation, Kauppinen \textit{et al.}\cite{Kauppinen_2019} also reported that the strain-adjusted GCN of the H-binding site could not explain variations in OH$^*$+H$^*$ adsorption energies on zirconia-supported Pt and Rh clusters. Our results extend this picture to unsupported finite Ni systems, showing that the O-site coordination provides a substantially more informative descriptor of the predicted adsorption-energy variations than the H-site coordination. Although GCN is not explicitly provided as an input feature, the GNN recovers these systematic structure--energy relationships from the local geometric and coordination information encoded in the graph.

Having characterized the DFT-validated NCs, we extend the analysis to larger systems as well, highlighting that these are prediction-only results. Nevertheless, this extension is motivated by the local nature of the framework: its applicability to larger particles depends primarily on whether their local surface environments are represented in the training domain, rather than on the total number of atoms. Across the investigated Ni sizes, the adsorption-site descriptor $\mathrm{GCN_O}$ spans approximately 1.8--10, with the DFT-validated systems covering a substantial portion of this range ($\sim$ 2.7--7.8, Figure \ref{fig:structural}d). The atomic GCN, which characterizes the local coordination environment of each surface Ni atom ($\mathrm{CN}_{\max}=12$), likewise shows that the larger NPs share many local environments with the training systems (Figure S13a). The bond-orientational order parameter $\bar{q}_6$\cite{Steinhardt_1983,Lechner_2008}, which describes local structural order, further indicates that the surface environments of the larger NPs often approach those of the Ni(111) and Ni(100) terminations (Figure S13b). Although these descriptors do not fully determine an adsorption-site environment or directly validate the prediction-only systems, their substantial overlap indicates that much of the larger-NP surface structure lies within a locally familiar regime. Predictions for the less familiar environments at the edges of these distributions should nevertheless be regarded as extrapolative.

The fitted slopes of $\mathrm{E}_{\mathrm{ads}}$ versus $\mathrm{GCN_O}$, shown for Ni$_{130}$ in Figure \ref{fig:structural}b, remain positive across the investigated NP sizes, with BH exhibiting the strongest coordination dependence for most systems (Figure \ref{fig:structural}e). Positive slopes of comparable magnitude have also been reported for atop OH$^*$ adsorption on late transition-metal NPs\cite{Hanselman_2023}. Consistent with stronger binding at lower $\mathrm{GCN_O}$, the lowest-energy BH and HH configurations tend to involve under-coordinated Ni atoms at corner/edge-like regions (examples for Ni$_{130}$, Ni$_{218}$, and Ni$_{686}$ in Figure S14)--reflecting the well-established tendency of under-coordinated metal sites on nanoparticles to bind adsorbates more strongly.\cite{callevallejo2014jpcl,Tereshchenko_2022} 

These predicted adsorption energies identify the regions of Ni NPs where the products of water dissociation are most strongly stabilized. However, favorable adsorption energies alone cannot establish whether these regions promote catalytic activity or contribute to surface poisoning, as both depend on the coverage and stability of reaction intermediates and the kinetics of subsequent elementary steps\cite{norskov2005,mohsenzadeh2014water}. An important advantage of the present framework is that it provides a systematic ranking of candidate O/H adsorption-site pairs, allowing favorable configurations to be identified across different regions of the Ni surface. These configurations could then be selected for targeted calculations of additional quantities, including adsorption free energies, coverage effects, and water-dissociation barriers, providing a more detailed picture of the processes governing catalytic activity. In this way, the proposed screening strategy could serve as a starting point for connecting local nanoparticle structure with catalytic performance.

\section{Conclusions}

We have developed an anchor-based GNN framework with an adsorbate-free representation for screening dissociative water adsorption on finite Ni nanosystems and low-index periodic surfaces. Exploiting the predominantly local nature of adsorption, the framework encodes each candidate O/H site pair using solely the Ni anchors that define its binding sites and their neighboring Ni atoms. Starting from the bare metal structure, the workflow automatically enumerates candidate site pairs, and separate GNNs predict their DFT-reference adsorption energies and whether the proposed binding sites would be retained under DFT relaxation. This representation eliminates the need to construct or relax adsorbate configurations during screening. Moreover, using MACE-relaxed bare Ni geometries as inputs reproduces the accuracy of the otherwise identical model based on DFT-relaxed geometries while avoiding DFT relaxation of the bare metal structures during screening.

The final M6-MACE model achieves an MAE of 0.032 eV on the internal validation set, outperforming the descriptor-based regression models. Across independent DFT tests on Ni$_{55}$, Ni$_{91}$, Ni$_{130}$, and larger Ni(111) and Ni(100) surface cells, it generally preserves relative energetic trends, although accuracy varies across systems (absolute-energy MAEs of 0.022--0.187 eV). On the same systems, the retention classifier provides a useful first-pass filter, with 93.6\% of predicted-retained configurations preserving their motifs and anchoring sites after DFT relaxation. Motivated by these results and the local nature of the representation, we also extend the analysis to prediction-only NPs containing up to $\sim$3,000 atoms. Structural descriptors indicate substantial overlap between their local surface environments and those represented in the training domain, although predictions for less familiar environments remain extrapolative. Across the Ni$_{55}$--Ni$_{3007}$ series, configurations passing the retention filter predominantly belong to three motif families: BH, HH, and TH. The BH and HH adsorption-energy distributions reach the most negative values, whereas TH tends to be less favorable. On Ni(111) and Ni(100), the predicted lowest-energy configurations instead exhibit HH and FF motifs, respectively, in agreement with literature results. Correlating the predicted adsorption energies of retained NP configurations with structural descriptors--such as the generalized coordination number--reveals that lower O-site coordination is associated with greater stabilization of the co-adsorbed OH$^*$+H$^*$ state, whereas H-site coordination shows a considerably weaker correlation. 

Finally, the presented framework provides a systematic route for mapping the stability of water-dissociation products across nanoparticle and low-index periodic surfaces and identifying promising configurations for further investigation. While a more complete assessment of their catalytic relevance would require additional calculations of adsorption free energies, coverage effects, and water-dissociation barriers, the proposed screening strategy can help prioritize the configurations for which these more demanding calculations are performed. Future extensions could also broaden the structural environments represented during training---for example, by including the more open Ni(110) and stepped surfaces. The anchor-based strategy could likewise be applied to other metals and adsorbate combinations by adapting the binding-site representations and retraining on suitable reference data, offering a route toward broader screening of co-adsorption energetics on complex catalytic nanostructures.

\begin{acknowledgement}
M.M., O.L., M.B.B., and B.E. acknowledge the HyPRO project, funded by GroenvermogenNL under Grant ID$\#$ NGF.1695.1695.012. All quantum-chemical calculations were performed on the Dutch national supercomputer Snellius, supported by SURF (grant number 2026.005).
\end{acknowledgement}

\section*{Data Availability}
All datasets and scripts associated with this work, including the DFT reference data and workflows for adsorption-configuration generation, GNN training, and prediction, are available at \url{https://github.com/mmonti00/adsorbate-free_GNN_Ni_NPs}.

\section*{Conflict of Interest}
The authors declare no conflict of interest.

\begin{suppinfo}

Additional computational details on adsorption-configuration generation, graph construction, GNN architectures, and training procedures; model selection and validation; motif-retention classification; comparisons with descriptor-based regression models; transferability to larger systems and computational screening costs; and structural determinants of adsorption-site stability.

\end{suppinfo}

\clearpage
\let\addcontentsline\siaddcontentsline
\newgeometry{left=28.35mm,right=28.35mm,top=16.6mm,bottom=16.6mm}
\singlespacing


\begin{center}
    \setstretch{2.0}
    {\Large\bfseries\sffamily Supplementary Information for:\\ "Adsorbate-Free Screening of Dissociative Water Adsorption on Nickel Nanoparticles via Graph Neural Networks"}
\end{center}

\vspace{0.5em}

\begin{center}
\setstretch{1.5}
\normalsize\sffamily
Marta Monti$^{*,\dagger}$, Alec Noppe$^{\dagger\ddagger}$, Olivier Lugier$^{\P}$, Maria B. Brands$^{\mathsection}$, and Bernd Ensing$^{*,\dagger,\ddagger}$
\end{center}


\begin{center}
    \setstretch{1.5}
    \textit{$^{\dagger}$Van ’t Hoff Institute for Molecular Sciences, University of Amsterdam, Science Park 904, 1098 XH Amsterdam, The Netherlands} \\
    \textit{$^{\ddagger}$AI4Science Laboratory, University of Amsterdam, Science Park 904, 1098 XH Amsterdam, The Netherlands} \\
    \textit{$^{\P}$Nano Hybrids B.V., Science Park 904, 1098 XH Amsterdam, The Netherlands}\\
    \textit{$^{\mathsection}$Elexel B.V., Science Park 408, 1098 XH Amsterdam, The Netherlands}\\
\end{center}

\vspace{0.3em}
\begin{center}
    \sffamily{E-mail:} m.monti@uva.nl; b.ensing@uva.nl
\end{center}

\vspace{2em}

\beginsupplement
\newpage
\tableofcontents
\newpage 

\makeatletter
\let\@startsection\acs@startsection
\makeatother


\section{Computational details}

\begin{figure} [H]
    \centering
    \includegraphics[width=1.0\linewidth]{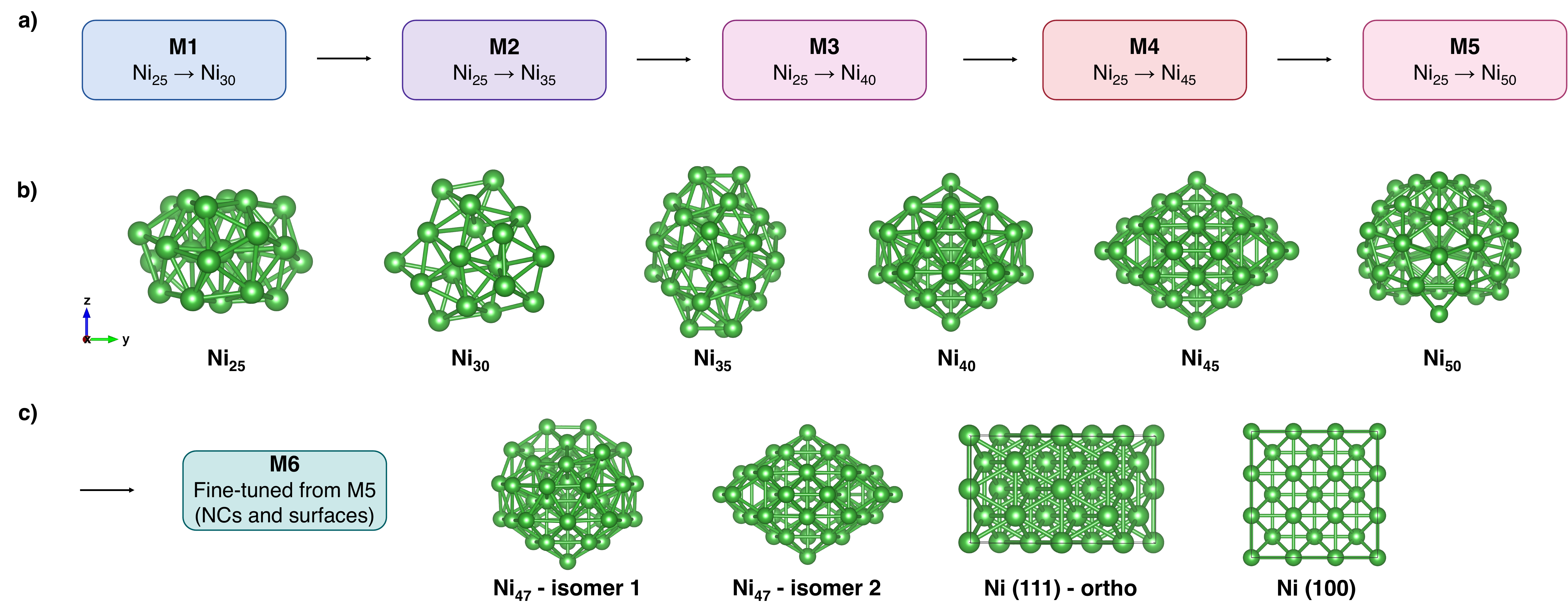}
    \caption{Overview of the progressive construction of the GNN model series for adsorption energy prediction. (a) Models M1--M5 were built by progressively expanding the nanocluster (NC) size coverage of the training set: M1 was trained on Ni$_{25}$ and Ni$_{30}$, and M2--M5 successively added Ni$_{35}$, Ni$_{40}$, Ni$_{45}$, and Ni$_{50}$ to the preceding training set. (b) Representative bare Ni NC structures in the corresponding datasets; the Cartesian reference axes shown for Ni$_{25}$ apply to all NC structures. (c) M6 was obtained by fine-tuning M5 on an expanded dataset that adds further NC configurations---two Ni$_{47}$ isomers---together with periodic Ni surfaces: orthorhombic Ni(111) and Ni(100), shown in top view.}
    \label{fig:M1_to_M6_struct}
\end{figure}

\subsection{Construction and filtering of candidate configurations}
\label{sec:si_site_generation}
Candidate OH$^*$+H$^*$ configurations were generated from ensembles of surface Ni atoms defining four nominal adsorption-site types: top (T), bridge (B), threefold hollow (H), and fourfold (F). A T site was associated with one surface Ni atom, while B, H, and F sites were associated with ensembles of two, three, and four surface Ni atoms, respectively. For Ni(111), the H category includes both the conventional fcc and hcp threefold hollow sites. For the nanoclusters (NCs), fourfold sites generally correspond to non-planar, folded fourfold motifs, whereas on Ni(100) the F category corresponds to the conventional fourfold hollow site. Representative examples of the four site types are shown in Figure 1b in the main text.

The geometrical reference position of a T site corresponded to the position of the associated Ni atom. For a B site, the reference position was the midpoint of the two Ni atoms, whereas the centroids of the corresponding three- and four-atom ensembles were used for H and F sites, respectively. An F candidate was retained only when its four surface Ni atoms formed a closed four-membered ring: each atom bonded to exactly two others in the set, so that the four atoms enclosed a single open square or rhombic hollow rather than a compact cluster. Four-atom groups that did not form such a ring---for example a three-atom face capped by an additional surface Ni atom---were excluded, because their centroid falls over a metal atom rather than over an exposed hollow. For each candidate site, an outward direction was determined from the local surface geometry. OH$^*$ and H$^*$ were placed at site-dependent distances above the corresponding reference position along this direction (of order $1.4$--$1.9$ and $1.0$--$1.6$ \AA, respectively). For OH$^*$, the hydroxyl H atom was placed $0.98$ \AA{} from the O atom along the same direction. The resulting coordinates were intended only to provide physically reasonable starting geometries representative of the selected adsorption motifs. An initial co-adsorption motif was denoted by two letters, with the first letter identifying the nominal site of OH$^*$ through its O atom and the second identifying the nominal site of the independently adsorbed H$^*$. For example, HT indicates OH$^*$ initially placed at a threefold hollow site and H$^*$ at a top site.

Not all possible combinations of the independently generated OH$^*$ and H$^*$ sites were converted into explicit adsorbed structures. Candidate pairs were discarded before the electronic-structure calculations when the two nominal sites represented the same or strongly overlapping region of the surface, since such combinations would place OH$^*$ and H$^*$ unrealistically close to one another and would not provide distinct co-adsorption configurations. The excluded combinations included identical adsorption sites and cases in which the anchor ensemble of one site was contained within that of the other. 

Each candidate structure surviving this filtering was subsequently relaxed according to the protocol reported in the \textit{Electronic-structure calculations} Section in the main text. During relaxation the adsorbates were free to move and could migrate away from their initial sites, so the adsorption motif of each configuration was determined only afterwards, from the relaxed geometry, as described in the following section.

\subsection{Anchor assignment and motif definition}
\label{sec:si_anchor_definition}
The motif of each relaxed configuration was assigned from the Ni atoms coordinating the two adsorbates, and could therefore differ from the initial motif used to generate the structure. Adsorbate migration was not itself used as a criterion for discarding a configuration.

For each relaxed structure, the hydroxyl H was identified as the H atom within $1.2$ \AA{} of the O atom, and the remaining H was treated as the independently adsorbed H$^*$. The coordinating Ni atoms were then determined by distance. For a standard assignment, the O anchors were the Ni atoms within $2.20$ \AA{} of O and the H anchors those within $2.00$ \AA{} of H$^*$, retaining at most the three shortest contacts in each case; the site type followed directly from the number of anchors, one, two, or three Ni atoms corresponding to T, B, or H. An F site was assigned when an adsorbate presented exactly four Ni contacts within an extended cutoff ($2.30$ \AA{} for O, $2.20$ \AA{} for H$^*$), of which at least two lay within the corresponding standard cutoff; in that case all four contacts were retained as anchors. The final motif was denoted by the pair of O- and H-site labels.

This assignment served two purposes. It fixed the motif label reported for each configuration, ensuring that the label reflected the relaxed adsorption geometry rather than the initial placement. It also provided the O- and H-anchor sets that constitute the sole encoding of the adsorbate positions used by the model: the GNN predicts the adsorption energy ($\mathrm{E}_{\mathrm{ads}}$) from the bare Ni geometry with these anchors flagged as node features, without any explicit adsorbate atoms (see Section \ref{sec:si_graph}).

\subsection{Post-relaxation screening and deduplication}
\label{sec:si_geo_filtering}
Before constructing the training datasets, the relaxed structures were screened to ensure that they represented distinct OH$^*$+H$^*$ co-adsorption configurations. Structures in which the dissociated species recombined to form molecular H$_2$O were removed, since they no longer belonged to the target OH$^*$+H$^*$ adsorption state. Duplicate configurations were identified from the O- and H-anchor sets assigned after relaxation (Section \ref{sec:si_anchor_definition}). Each configuration was represented by a site signature combining these anchor sets with the identity of the bare Ni structure. Two configurations were considered duplicates when they referred to the same bare NC structure or structural isomer and had identical O- and H-anchor sets, in which case only the first occurrence was retained. Distinct NC isomers or conformations were treated separately, even when they produced formally equivalent anchor patterns. This procedure removed configurations that had converged to the same relaxed adsorption environment despite originating from different initial motifs or candidate-generation steps. The resulting cleaned structures and their reference adsorption energies were used to construct the datasets employed for training and validation.

\subsection{Graph construction and node features for the energy model}
\label{sec:si_graph}

Each Ni+OH$^*$+H$^*$ configuration was mapped to a graph built from the geometry of the corresponding \emph{bare} Ni structure. Let $\mathbf{R}_1, \dots, \mathbf{R}_N \in \mathbb{R}^3$ denote the positions of the $N$ Ni atoms in the bare structure, and let $d_{ij} = \lVert \mathbf{R}_i - \mathbf{R}_j \rVert$. Let $\mathcal{A}_\mathrm{O}, \mathcal{A}_\mathrm{H} \subseteq \{1, \dots, N\}$ be the indices of the O- and H-anchor atoms identified from the relaxed structure (Section~\ref{sec:si_anchor_definition}), and $\mathcal{A} = \mathcal{A}_\mathrm{O} \cup \mathcal{A}_\mathrm{H}$. 

Because the structural response to adsorption is predominantly local, the graph was restricted to Ni atoms near the adsorption site. The node set and directed edge set were 
\begin{equation}
\begin{aligned}
  \mathcal{V} &= \bigl\{\, i \in \{1, \dots, N\}: \min_{a \in \mathcal{A}} d_{ia} \le r_\mathrm{loc} \,\bigr\}, \\
  \mathcal{E} &= \bigl\{\, (i,j) \in \mathcal{V} \times \mathcal{V} : i \neq j,\; d_{ij} \le r_\mathrm{edge} \,\bigr\},
\end{aligned}
\label{eq:graph_sets}
\end{equation}
with $r_\mathrm{loc} = 5.5$~\AA{} and $r_\mathrm{edge} = 4.0$~\AA{}. Since $d_{ij} = d_{ji}$, every neighboring pair contributes an edge in each direction.

Each node $i \in \mathcal{V}$ carried an eight-dimensional feature vector
\begin{equation}
\mathbf{x}_i =
\bigl(\,1,\;
\mathbf{1}[i\in\mathcal{A}_\mathrm{O}],\;
\mathbf{1}[i\in\mathcal{A}_\mathrm{H}],\;
\mathbf{1}[i\in\mathcal{A}],\;
\lVert\mathbf{R}_i-\mathbf{c}_\mathrm{O}\rVert,\;
\lVert\mathbf{R}_i-\mathbf{c}_\mathrm{H}\rVert,\;
\lVert\mathbf{R}_i-\mathbf{c}_\mathcal{A}\rVert,\;
n_i\,\bigr),
\label{eq:node_features}
\end{equation}
where the first entry is a constant Ni-identity feature, the indicators flag O-, H-, and any-anchor membership, and $\mathbf{c}_\mathrm{O}$, $\mathbf{c}_\mathrm{H}$, $\mathbf{c}_\mathcal{A}$ are the centroids of the atoms in $\mathcal{A}_\mathrm{O}$, $\mathcal{A}_\mathrm{H}$, and $\mathcal{A}$. The final entry,
\begin{equation}
  n_i = \bigl\lvert\{\, j \in \mathcal{V} : j \neq i,\; d_{ij} \le r_\mathrm{edge} \,\}\bigr\rvert,
\label{eq:neighbor_count}
\end{equation}
counts the Ni neighbors of $i$ within the local node set $\mathcal{V}$ and therefore equals the degree of node $i$ in the local graph. All features depend only on distances and are therefore invariant under the Euclidean group E(3) of translations, rotations, and reflections.

Each edge $(i,j) \in \mathcal{E}$ carried its distance expanded in a basis of $K = 16$ Gaussian radial functions~\cite{schutt2018schnet}, 
\begin{equation}
[\mathbf{e}_{ij}]_k = \exp\!\bigl[-\gamma\,(d_{ij}-\mu_k)^2\bigr],
\qquad k=1,\dots,K,
\label{eq:rbf}
\end{equation}
with centers $\mu_k$ evenly spaced over $[0,5]$~\AA{} and $\gamma=10$~\AA$^{-2}$.

\subsection{Network architecture for the energy model}
\label{sec:si_architecture}
The model applies $L=3$ edge-conditioned message-passing layers with hidden dimension $d=128$, initialized from $\mathbf{h}_i^{(0)}=\mathbf{x}_i$. The architecture follows the neural message-passing framework\cite{gilmer2017mpnn}, with the edge-gating construction inspired by the EGNN (edge-inference) formulation of Satorras \emph{et al.}\cite{satorras2021egnn}. At layer $\ell$, a message is constructed for each edge from the representations of the two connected nodes and the corresponding edge embedding; the messages are then gated by a learned scalar coefficient and aggregated over neighboring atoms:
\begin{align}
\mathbf{m}_{ij}^{(\ell)} &= \phi_\mathrm{e}^{(\ell)}\bigl(
  \mathbf{h}_i^{(\ell)} \,\Vert\, \mathbf{h}_j^{(\ell)} \,\Vert\, \mathbf{e}_{ij}\bigr),
  \label{eq:message}\\
\alpha_{ij}^{(\ell)} &= \sigma\bigl(\phi_\alpha^{(\ell)}(\mathbf{m}_{ij}^{(\ell)})\bigr),
  \label{eq:attention}\\
\mathbf{m}_i^{(\ell)} &= \sum_{j \in \mathcal{N}(i)} \alpha_{ij}^{(\ell)}\,\mathbf{m}_{ij}^{(\ell)},
  \label{eq:aggregate}\\
\mathbf{h}_i^{(\ell+1)} &= \mathrm{ReLU}\Bigl(
  \mathbf{P}^{(\ell)}\mathbf{h}_i^{(\ell)}
  + \phi_\mathrm{n}^{(\ell)}\bigl(\mathbf{h}_i^{(\ell)} \,\Vert\, \mathbf{m}_i^{(\ell)}\bigr)\Bigr),
  \label{eq:update}
\end{align}
where $\mathcal{N}(i) = \{\, j : (i,j) \in \mathcal{E} \,\}$, $\Vert$ denotes concatenation, $\sigma$ is the logistic sigmoid, $\alpha_{ij}^{(\ell)}$ is the learned scalar edge-gating coefficient, and $\mathbf{P}^{(\ell)}$ is a linear residual projection (the identity when input and hidden dimensions coincide). The message and update functions $\phi_\mathrm{e}^{(\ell)}$ and $\phi_\mathrm{n}^{(\ell)}$ are two-layer perceptrons with ReLU activations~\cite{nair2010relu,glorot2011relu}, each mapping its concatenated input to $\mathbb{R}^d$, and the gate function $\phi_\alpha^{(\ell)} : \mathbb{R}^d \to \mathbb{R}$ is affine.

Let $\mathbf{H}_i=\mathbf{h}_i^{(L)}$ denote the final node representations. These were combined through a custom mixed global--local pooling scheme producing three graph-level vectors. A global representation was obtained as the mean of the final node features over the subgraph,
\begin{equation}
\mathbf{g}_\mathrm{global} = \frac{1}{|\mathcal{V}|}\sum_{i\in\mathcal{V}}\mathbf{H}_i,
\label{eq:pool_global}
\end{equation}
while two anchor-centered representations were obtained by distance-weighted mean pooling around the O and H adsorption sites,
\begin{align}
&w_i^{\mathrm{O}} = \exp\!\bigl(-\lVert\mathbf{R}_i-\mathbf{c}_\mathrm{O}\rVert/\sigma_\mathrm{p}\bigr)+\beta\,\mathbf{1}[i\in\mathcal{A}_\mathrm{O}],
\qquad
\mathbf{g}_\mathrm{O}=\frac{\sum_{i}w_i^{\mathrm{O}}\mathbf{H}_i}{\sum_{i}w_i^{\mathrm{O}}},
\label{eq:pool_O}\\[2pt]
&w_i^{\mathrm{H}} = \exp\!\bigl(-\lVert\mathbf{R}_i-\mathbf{c}_\mathrm{H}\rVert/\sigma_\mathrm{p}\bigr)+\beta\,\mathbf{1}[i\in\mathcal{A}_\mathrm{H}],
\qquad
\mathbf{g}_\mathrm{H}=\frac{\sum_{i}w_i^{\mathrm{H}}\mathbf{H}_i}{\sum_{i}w_i^{\mathrm{H}}}.
\label{eq:pool_H}
\end{align}
with $\sigma_\mathrm{p}=1.5$ \AA{} and $\beta=2$. The pooling weights depend only on the fixed geometric features of Eq. \ref{eq:node_features}, so they emphasize atoms near the O and H adsorption sites and boost the anchor atoms themselves. The three vectors were concatenated to obtain the final graph representation, which was mapped to the predicted adsorption energy through a fully connected neural network:
\begin{equation}
\hat{\mathrm{E}}_\mathrm{ads} =
\phi_\mathrm{out}\!\bigl(\,\mathbf{g}_\mathrm{global}\,\Vert\,\mathbf{g}_\mathrm{O}\,\Vert\,\mathbf{g}_\mathrm{H}\,\bigr),
\label{eq:energy_head}
\end{equation}
where $\phi_\mathrm{out}$ is a two-layer perceptron ($3d\rightarrow64\rightarrow1$) with a ReLU activation between the layers. The complete GNN contains $282,756$ trainable parameters.

\subsection{Training of the M1--M5 energy models}
\label{sec:si_training}
Models M1--M5 (see Figure \ref{fig:M1_to_M6_struct}) were trained by minimizing the mean-squared error between predicted and DFT adsorption energies,
\begin{equation}
\mathcal{L} = \frac{1}{|\mathcal{B}|}\sum_{n\in\mathcal{B}}
\bigl(\hat{\mathrm{E}}_{\mathrm{ads},n}-\mathrm{E}_{\mathrm{ads},n}\bigr)^2,
\label{eq:loss}
\end{equation}
over mini-batches $\mathcal{B}$. Optimization used Adam\cite{kingma2015adam} with a learning rate of $8\times10^{-4}$, a weight decay of $1\times10^{-6}$, and a batch size of $32$, for a maximum of $200$ epochs. Training was stopped early when the validation MAE did not improve for $25$ consecutive epochs, and the parameters giving the lowest validation MAE were retained. Each dataset was partitioned into $80\%$ training and $20\%$ validation subsets by stratified sampling on the joint label of NC size and final OH$^*$+H$^*$ motif. A single random seed ($32$) controlled the dataset partitioning, parameter initialization, and mini-batch ordering.

\subsection{Fine-tuning of the M6 energy model}
\label{sec:si_finetuning}
The final M6 model was obtained by fine-tuning the M5 checkpoint on an extended dataset combining NC and periodic Ni(111) and Ni(100) surface configurations as schematized in Figure \ref{fig:M1_to_M6_struct}. All parameters of M5---message-passing layers, pooling, and energy head---were used to initialize M6, and all were left trainable; the architecture, hidden dimension, node and edge features, anchor definitions, and H$_2$O-referenced target convention were kept identical to those of M1--M5. Graph construction, node features (Eq. \ref{eq:node_features}), the radial basis (Eq. \ref{eq:rbf}), the edge-conditioned message passing (Eqs. \ref{eq:message}--\ref{eq:update}), the mixed global--local pooling (Eqs. \ref{eq:pool_global}--\ref{eq:pool_H}), and the energy head (Eq. \ref{eq:energy_head}) were all retained unchanged.

For the periodic surface structures, periodicity was treated explicitly during graph construction. Minimum-image distances, as implemented in the Atomic Simulation Environment\cite{larsen2017ase}, were used to resolve the periodic images of the O- and H-anchor sites. Periodic replicas of the Ni atoms were then generated around the adsorption region, and the local graph was constructed from all image atoms within the prescribed cutoff. Coordination numbers, edge distances, and distances to the anchor centroids were subsequently evaluated using Cartesian distances within this finite periodic-image graph. This construction ensures that equivalent local adsorption environments are represented consistently irrespective of their position relative to the periodic-cell boundaries. The NC graphs retained the local-environment and edge cutoffs of $5.5$ and $4.0$ \AA{} used for M1--M5, whereas values of $6.0$ and $4.5$ \AA{} were used for the surfaces.

Because the number of surface configurations was much smaller than that of the NC configurations, each sample was assigned a regression weight $w_n$ in a weighted mean-squared-error loss,
\begin{equation}
\mathcal{L} =
\frac{\sum_{n\in\mathcal{B}} w_n\,\bigl(\hat{\mathrm{E}}_{\mathrm{ads},n}-\mathrm{E}_{\mathrm{ads},n}\bigr)^2}
{\sum_{n\in\mathcal{B}} w_n},
\label{eq:weighted_loss}
\end{equation}
with $w_n=1$ for NC and $w_n=4$ for surface configurations. Model selection and early stopping, however, were based on the \emph{unweighted} validation MAE, so that the up-weighting of the surface structures in the training loss did not dominate checkpoint selection or mask a degradation of the NC predictions.

Fine-tuning used the Adam optimizer\cite{kingma2015adam} with a learning rate of $3\times10^{-4}$, a weight decay of $1\times10^{-6}$, and a batch size of $32$, for a maximum of $100$ epochs with an early-stopping patience of $25$ epochs on the unweighted validation MAE. The NC and surface datasets were each partitioned independently into $80\%$ training and $20\%$ validation subsets by stratified sampling on the joint label of structural group and final OH$^*$+H$^*$ motif within each domain, and the resulting training and validation subsets were subsequently combined. A single random seed ($32$) controlled the partitioning and mini-batch ordering. Because the model input depends only on the bare Ni geometry and the anchor sets, symmetry-equivalent adsorption sites of the same structure are represented by identical graphs and can be assigned to different subsets; this applies to approximately 20\% of the NC and more than 90\% of the surface validation configurations. The internal validation therefore primarily assesses predictions for new adsorption sites on structures represented in training, whereas transferability to unseen structures is evaluated on the external test systems.

\section{Architecture assessment and progressive training-set expansion (energy model)}
\label{sec:ablation_overview}
To assess the influence of the main architectural and training choices on the prediction of adsorption energies, we performed a series of controlled tests using M5 (Ni$_{25}$ $\rightarrow$ Ni$_{50}$, Figure \ref{fig:M1_to_M6_struct}) as the reference model. Unless otherwise stated, all model variants were trained using the same fixed training/validation partition, graph representation, message-passing architecture, loss function, and optimization settings, such that only the investigated component was varied. Model performance was evaluated on the validation set using the mean absolute error (MAE), root-mean-square error (RMSE), and coefficient of determination (R$^2$).

\subsection{Pooling strategy}
\label{sec:ablation_pooling}

We first examined how the construction of the graph-level representation affects the prediction of the adsorption energy. Three pooling strategies were considered: global mean pooling over the complete local Ni environment, separate adsorbate-centered pooling around the O and H anchoring sites, and a mixed representation obtained by combining the global and adsorbate-centered embeddings. 

To isolate the effect of the pooling operation, all three models were trained using the same fixed training/validation partition, random seed, message-passing architecture, and optimization settings. In addition, a dedicated PyTorch random-number generator, initialized with the same seed for each model, was used for the shuffled training DataLoader. This ensured that the three models were exposed to the same sequence of mini-batches during training.

\begin{figure} [H]
    \centering
    \includegraphics[width=1.0\linewidth]{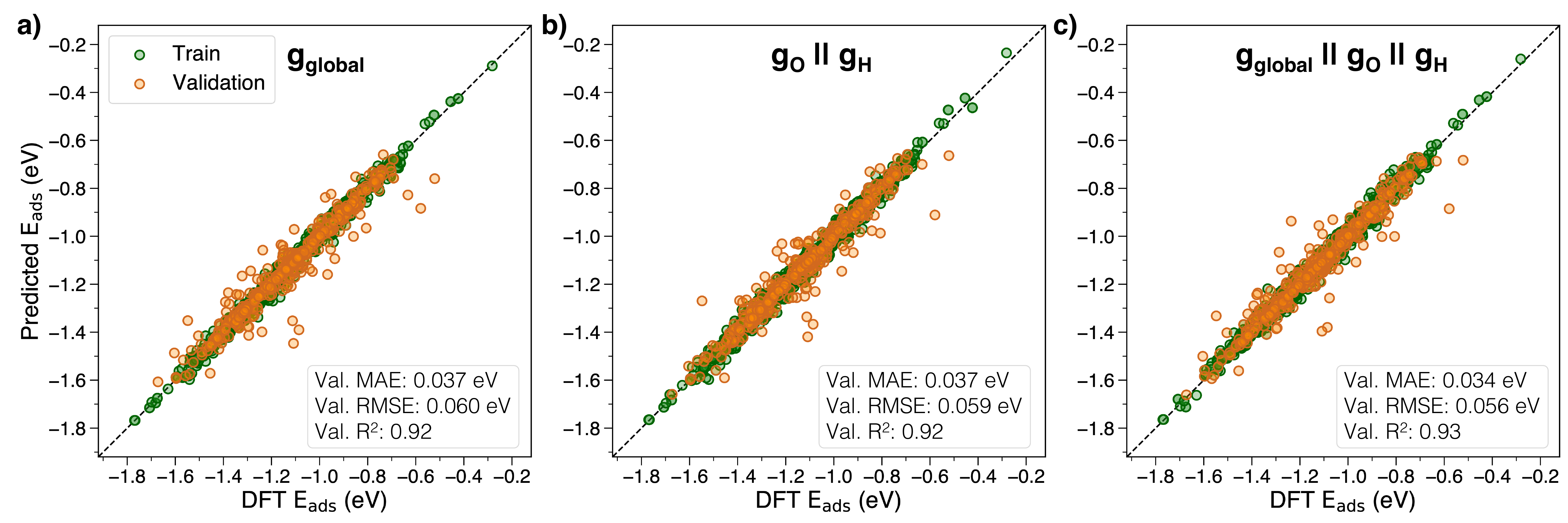}
    \caption{Pooling-strategy ablation for the adsorption energy GNN. Parity plots compare DFT and predicted adsorption energies (E$_{\mathrm{ads}}$) for models employing (a) global mean pooling, (b) separate O- and H-focused local pooling, and (c) concatenated global and local representations. All models were trained using the same fixed train/validation partition, random seed, message-passing architecture, and optimization settings. Green and orange markers denote training and validation structures, respectively, while the dashed line indicates ideal agreement between predicted and DFT energies. The reported mean absolute error (MAE), root-mean-square error (RMSE), and coefficient of determination (R$^2$) refer to the validation set.}
    \label{fig:parity_pooling}
\end{figure}

The global-pooling model yielded a validation MAE of 0.037 eV, an RMSE of 0.060 eV, and an R$^2$ of 0.92. Using only the separate O- and H-centered representations produced very similar performance, with a validation MAE of 0.037 eV, an RMSE of 0.059 eV, and an R$^2$ of 0.92. The best performance was obtained by combining the global and adsorbate-centered representations, resulting in a validation MAE of 0.034 eV, an RMSE of 0.056 eV, and an R$^2$ of 0.93 (Figure \ref{fig:parity_pooling}). 

These results suggest that the adsorbate-centered representations benefit from being complemented by an unweighted representation of the complete local Ni subgraph. The mixed pooling strategy therefore captures both the immediate environments of the O and H anchoring sites and the surrounding Ni environment within the selected cutoff radius, and was retained for the subsequent models.

\subsection{Model capacity}
\label{sec:ablation_capacity}

\begin{figure} [H]
    \centering
    \includegraphics[width=1.0\linewidth]{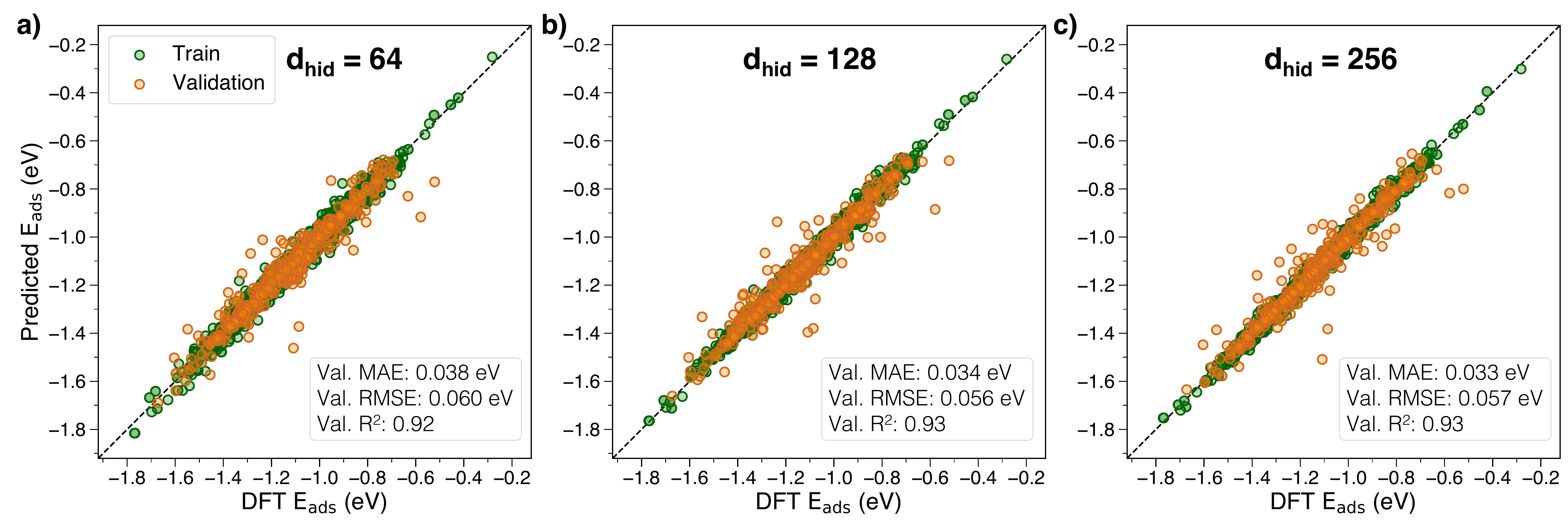}
    \caption{Effect of model capacity on adsorption energy prediction. Parity plots compare DFT reference and predicted adsorption energies (E$_{\mathrm{ads}}$) for models with hidden dimensions (d$_{hid}$) of (a) 64, (b) 128, and (c) 256. All models were trained using the same fixed train/validation partition, random seed, message-passing architecture, and optimization settings. Green and orange markers denote training and validation structures, respectively, while the dashed line indicates ideal agreement between predicted and DFT energies. The reported mean absolute error (MAE), root-mean-square error (RMSE), and coefficient of determination (R$^2$) values refer to the validation set.}
    \label{fig:parity_hid_dim}
\end{figure}

The effect of model capacity was investigated by varying the hidden dimension, $d_{\mathrm{hid}}$, while retaining the mixed global--local pooling strategy described in Subsection \ref{sec:ablation_pooling}. Three models were considered: (i) $d_{\mathrm{hid}}=64$ (80,004 trainable parameters), (ii) $d_{\mathrm{hid}}=128$ (282,756 parameters), and (iii) $d_{\mathrm{hid}}=256$ (1,056,900 parameters).
All models were trained using the same fixed training/validation partition, random seed, mini-batch sequence, message-passing architecture, and optimization settings. Resulting parity plots are shown in Figure \ref{fig:parity_hid_dim}.

Increasing the hidden dimension from 64 to 128 improved the validation performance. The validation MAE decreased from 0.038 to 0.034 eV, while the RMSE decreased from 0.060 to 0.056 eV and the R$^2$ increased from 0.92 to 0.93. A further increase to $d_{\mathrm{hid}}=256$ produced only negligible changes in the validation metrics (MAE = 0.033 eV, RMSE = 0.057 eV, and R$^2$ = 0.93). Thus, most of the improvement associated with increasing model capacity was already achieved at $d_{\mathrm{hid}}=128$. Since increasing the hidden dimension from 128 to 256 significantly raised the number of trainable parameters while only reducing the validation MAE by 0.001 eV, the model with $d_{\mathrm{hid}}=128$ was retained as the final architecture, providing a favorable compromise between predictive performance and model complexity.

\subsection{Random-seed sensitivity}
\label{sec:random_seed_sensitivity}

The sensitivity of the training procedure to stochastic initialization was evaluated by repeating the training of the M5 model using ten different random seeds. The training/validation partition, model architecture, and optimization settings were kept fixed. In this analysis, the random seed controlled both the initialization of the model parameters and the shuffled ordering of the training mini-batches. Thus, the test probes the overall variability associated with the stochastic training trajectory.

\begin{figure} [H]
    \centering
    \includegraphics[width=1.0\linewidth]{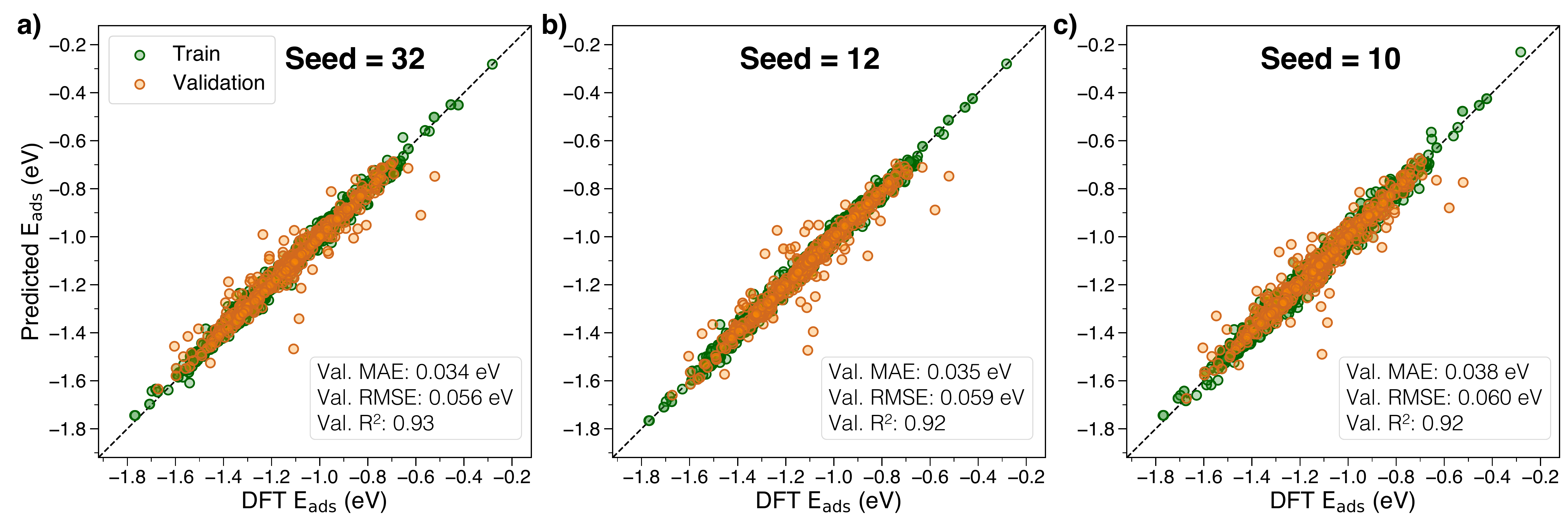}
    \caption{Effect of the random seed on M5 adsorption energy prediction. Parity plots compare DFT reference and predicted adsorption energies (E$_{\mathrm{ads}}$) for representative runs corresponding to (a) seed 32, (b) seed 12, and (c) seed 10. These runs represent the lowest, median, and highest validation mean absolute error (MAE), respectively, among the ten random seeds tested (see Table \ref{tab:M5_seed_sensitivity}). All models were trained using the same fixed train/validation partition, message-passing architecture, and optimization settings. Green and orange markers denote training and validation structures, respectively, while the dashed line indicates ideal agreement between predicted and DFT energies. The reported mean absolute error (MAE), root-mean-square error (RMSE), and coefficient of determination (R$^2$) refer to the validation set.
}
    \label{fig:parity_rand_seed_M5}
\end{figure}

Figure \ref{fig:parity_rand_seed_M5} shows representative parity plots for the runs with the lowest, median, and highest validation MAE. All ten runs achieved excellent predictive accuracy (Table \ref{tab:M5_seed_sensitivity}), with validation MAEs ranging from 0.034 to 0.038 eV and a mean value of $(0.036 \pm 0.002)$ eV. The validation RMSEs ranged from 0.055 to 0.062 eV, with a mean value of $(0.059 \pm 0.002)$ eV, while the corresponding R$^2$ values ranged from 0.91 to 0.93. The consistently low errors and narrow variation across seeds demonstrate both the high predictive quality of M5 and its robustness with respect to the stochastic training trajectory.

\begin{table}[H]
    \centering
    \caption{Sensitivity of the M5 adsorption energy model to the random seed. The reported mean absolute error (MAE), root-mean-square error (RMSE), and coefficient of determination (R$^2$) were calculated on the fixed validation set. The final row reports the mean and sample standard deviation (SD) calculated across the ten random-seed runs.}
    \label{tab:M5_seed_sensitivity}

    \renewcommand{\arraystretch}{1.35}
    \setlength{\tabcolsep}{12pt}

    \begin{tabular}{|c|c|c|c|}
        \hline
        \textbf{Seed} &
        \textbf{Val. MAE (eV)} &
        \textbf{Val. RMSE (eV)} &
        \textbf{Val. R$^2$} \\
        \hline
        2   & 0.034 & 0.059 & 0.92 \\
        \hline
        10  & 0.038 & 0.060 & 0.92 \\
        \hline
        12  & 0.035 & 0.059 & 0.92 \\
        \hline
        20  & 0.037 & 0.061 & 0.92 \\
        \hline
        22  & 0.035 & 0.055 & 0.93 \\
        \hline
        25  & 0.035 & 0.057 & 0.93 \\
        \hline
        32  & 0.034 & 0.056 & 0.93 \\
        \hline
        35  & 0.037 & 0.060 & 0.92 \\
        \hline
        37  & 0.036 & 0.062 & 0.91  \\
        \hline
        42  & 0.037 & 0.057 & 0.93  \\
        \hline\hline
        \textbf{Mean $\pm$ SD} &
        $0.036 \pm 0.002$ &
        $0.059 \pm 0.002$ &
        $0.92 \pm 0.01$ \\
        \hline
    \end{tabular}

    \vspace{1.0cm}
\end{table}

\subsection{Progressive expansion of the nanocluster training set}
\label{sec:M1_to_M5}
We present the parity plots of the M1--M5 model series, illustrating how the prediction quality evolves as the NC-size coverage of the dataset is progressively extended. All models share the final architecture and optimization protocol; the dataset size and NC-size coverage increase across the series. The M1 dataset comprised approximately 1550 configurations of Ni$_{25}$ and Ni$_{30}$. For M2--M5, the NC-size coverage was progressively expanded by including 30--80 configurations of Ni$_{35}$, Ni$_{40}$, Ni$_{45}$, and Ni$_{50}$, respectively, and each model was retrained from scratch. At each stage, an approximately 80/20 train/validation split was generated using the same seed and stratification by NC size and adsorption motif. We note that the M5 results shown in this section correspond to a retraining of the finalized architecture using a newly generated train/validation split and therefore differ slightly from those reported in the architecture-ablation analysis, where a common split was retained to enable a controlled comparison across model variants. 

\begin{figure} [H]
    \centering
    \includegraphics[width=1.0\linewidth]{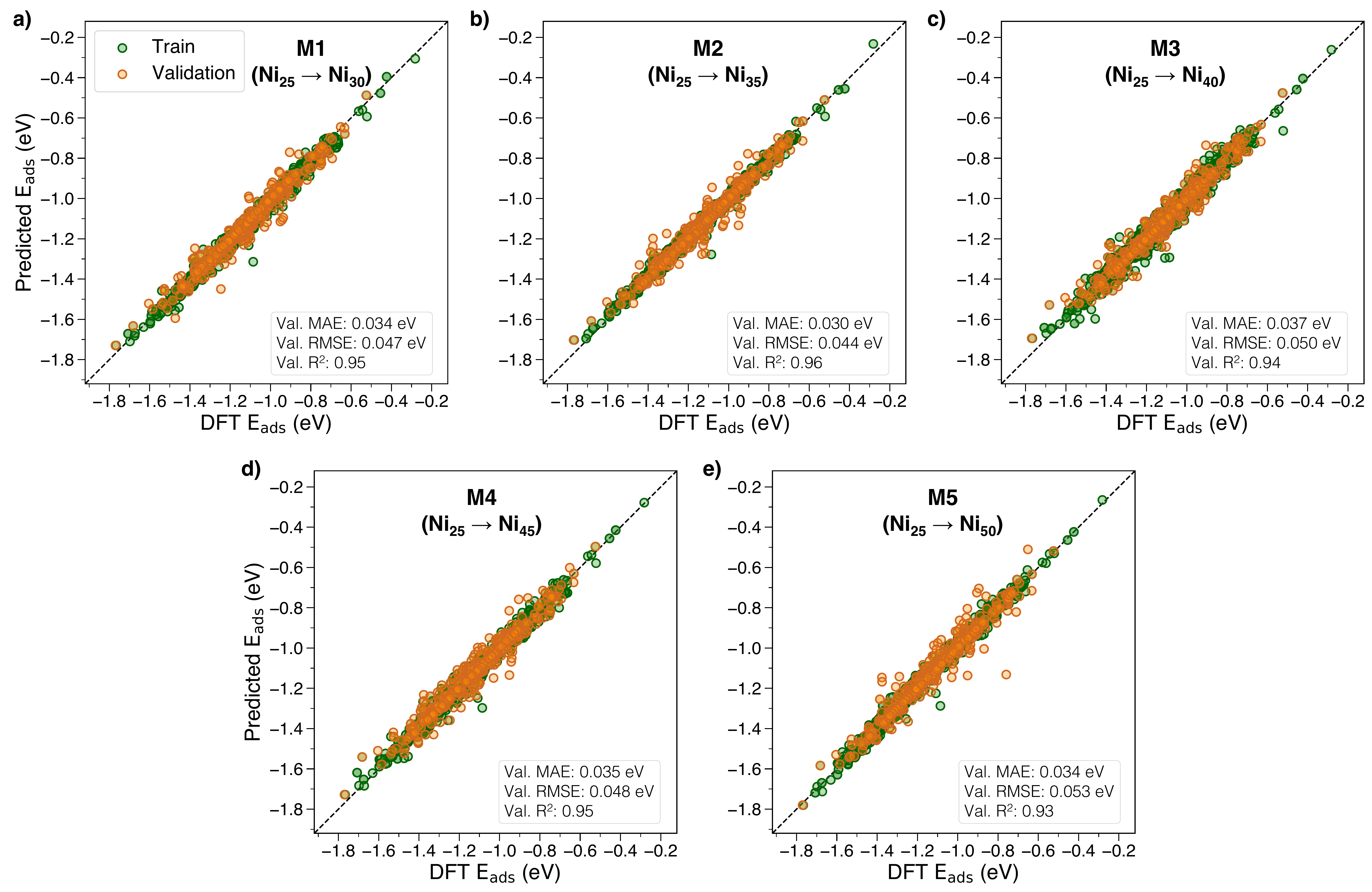}
    \caption{Progressive expansion of the nanocluster (NC) training set for the adsorption energy GNN. Parity plots compare DFT reference and predicted adsorption energies (E$_{\mathrm{ads}}$) for models (a) M1, trained on Ni$_{25}$ and Ni$_{30}$ NCs, and \textbf{(b--e)} M2–M5, obtained by progressively adding Ni$_{35}$, Ni$_{40}$, Ni$_{45}$, and Ni$_{50}$ structures, respectively. All models employed the same graph architecture, random seed, hyperparameters, and training protocol. Green and orange markers denote training and validation structures, respectively, while the dashed line indicates ideal agreement between predicted and DFT energies. The reported mean absolute error (MAE), root-mean-square error (RMSE), and coefficient of determination (R$^2$) refer to the validation set.}
    \label{fig:parity_M1_to_M5}
\end{figure}

Across the M1--M5 series, all models maintained consistently high predictive accuracy despite the progressive increase in NC-size coverage and structural diversity of the dataset (Figure \ref{fig:parity_M1_to_M5}). The validation MAEs remained within a narrow range of 0.030--0.037 eV, while the RMSEs ranged from 0.044 to 0.053 eV and the corresponding R$^2$ values from 0.93 to 0.96. No systematic deterioration in performance was observed as additional NC sizes were incorporated. In particular, M5, which spans the broadest range from Ni$_{25}$ to Ni$_{50}$, retained a validation MAE of 0.034 eV, comparable to that of M1. These results indicate that progressively extending the dataset with a relatively small number of configurations for each additional NC size preserves the predictive accuracy of the model while increasing the range of local Ni environments represented during training. 

\section{Motif-retention model: construction and validation}
\label{sec:si_retention}

\subsection{Retention target and dataset construction}
\label{sec:si_motif_data}
The motif-retention model was trained to predict whether the adsorption site assigned to an initially generated OH$^*$+H$^*$ configuration is preserved upon structural DFT relaxation. The input and target labels were therefore constructed from paired pre- and post-relaxation structures. A configuration was assigned to the retained class ($y=1$) only when both the final adsorption motif and the complete O- and H-anchor sets were identical to those of the initial structure. All remaining configurations were assigned to the migrated class ($y=0$). The latter therefore includes both configurations that preserve the same nominal motif while moving to a different set of Ni anchor atoms and configurations that relax to a different adsorption motif. This strict definition was adopted because the classifier is intended to identify candidate configurations for which the initially encoded adsorption environment is expected to remain unchanged during structural relaxation. Unlike the adsorption-energy dataset (see Section \ref{sec:si_geo_filtering}), no deduplication was performed for the retention dataset, as distinct initial configurations may converge to the same relaxed adsorption basin. Retaining these cases preserves information about the migration of different initial binding-site assignments during relaxation. The resulting retention dataset comprised 5,926 configurations (5,658 nanocluster and 268 periodic-surface configurations).

\subsection{Graph representation and network architecture}
\label{sec:si_motif_graph}

Each candidate configuration was represented by a local graph constructed from the bare Ni geometry and centered on the O- and H-anchor sets identified in the \emph{initial} structure. The node and edge representations followed those used for the adsorption-energy model (Section \ref{sec:si_graph}), with the anchor-membership flags and anchor-centered distances evaluated relative to the initial adsorption sites. For NC configurations, the node set comprised Ni atoms lying within 5.5 \AA{} of at least one O or H anchor, and directed edges were generated between retained Ni atoms separated by at most 4.0 \AA{}. For the periodic surface configurations, the corresponding local-environment and edge cutoffs were 6.0 and 4.5 \AA{}, respectively. For periodic surface configurations, minimum-image distances were used to identify the adsorption anchors and resolve their periodic images. Explicit replicas of the Ni atoms were subsequently generated around the adsorption sites, and the local graph was constructed from the resulting finite periodic-image environment.

The motif-retention model employed the same three-layer edge-gated message-passing encoder defined for the adsorption-energy model in Eqs. \ref{eq:message}--\ref{eq:update}, but with hidden dimension $d=64$. The graph-level representation was constructed using the same O- and H-centered distance-weighted pooling operations defined in Eqs. \ref{eq:pool_O} and \ref{eq:pool_H}. For the retention model, however, the anchor sets and corresponding centroids were those identified from the \emph{initial} candidate geometry, and the pooling parameters were set to $\sigma_\mathrm{p}=1.0$ \AA{} and $\beta=1.0$. In contrast to the final adsorption-energy model, no global mean-pooled representation was included. The resulting graph representation was therefore
\begin{equation}
\mathbf{g}_{\mathrm{ret}} =
\,\mathbf{g}_{\mathrm{O}} \Vert \mathbf{g}_{\mathrm{H}}\,.
\label{eq:retention_pool}
\end{equation}

A single message-passing encoder was shared between NC and periodic-surface configurations. After pooling, $\mathbf{g}_{\mathrm{ret}}$ was passed to one of two domain-specific binary classification heads, one for NCs and one for slabs. Each head consisted of a two-layer fully connected network ($2d\rightarrow64\rightarrow1$) with a ReLU activation and a dropout probability of 0.1. The structural domain was used only to select the corresponding output head and was not provided as a node or graph feature. Each head returned a scalar logit $z$, which was transformed through the logistic sigmoid,
\begin{equation}
s_{\mathrm{ret}} = \frac{1}{1+\exp(-z)},
\label{eq:retention_score}
\end{equation}
to obtain the retention score $s_{\mathrm{ret}}$.

\subsection{Training protocol}
\label{sec:si_motif_train}
The motif-retention classifier was trained using the unweighted binary cross-entropy (BCE) loss with logits. For a configuration $n$ with binary target $y_n$ and predicted logit $z_n$, the loss is

\begin{equation}
\mathcal{L}_{\mathrm{BCE}}
=
-\frac{1}{|\mathcal{B}|}
\sum_{n\in\mathcal{B}}
\left[
y_n\log\sigma(z_n)
+
(1-y_n)\log\left(1-\sigma(z_n)\right)
\right],
\end{equation}
where $\mathcal{B}$ denotes the mini-batch and $\sigma$ is the logistic sigmoid. No class-dependent weighting was applied. Because the number of NC configurations was substantially larger than that of the periodic-surface configurations, the NC and slab BCE losses were evaluated independently for the samples of each domain present in a mini-batch and combined using fixed domain weights of 0.80 and 0.20, respectively, before backpropagation. This weighting retained an explicit contribution from the surface domain while preventing the comparatively small slab subset from exerting disproportionate influence on the shared encoder.

The NC and surface datasets were partitioned independently into training and validation subsets, using approximately 80$\%$ and 20$\%$ of the data, respectively. To reduce information leakage between closely related configurations, the partitioning was performed at the level of the initial adsorption site. Configurations belonging to the same structural group and sharing the same initial motif and O- and H-anchor sets were therefore assigned to the same subset. The resulting partition was stored and kept fixed throughout all architecture, hyperparameter, and random-seed tests. Since symmetry-equivalent sites carry different anchor indices, they can nevertheless be assigned to different subsets while being represented by identical graphs; this is the case for approximately 22\% of the NC and more than 80\% of the surface validation configurations. The validation metrics therefore primarily assess new adsorption sites on structures represented in training, whereas transferability to unseen structures is assessed on the external test systems.

Optimization was performed using Adam with a learning rate of $8\times10^{-4}$, a weight decay of $1\times10^{-6}$, and a mini-batch size of 32. Training was carried out for a maximum of 200 epochs, with early stopping after 25 consecutive epochs without improvement in the domain-macro validation BCE, defined as the mean of the unweighted NC and slab validation BCE losses. The checkpoint with the lowest domain-macro validation BCE was retained, such that model selection reflected the performance of both structural domains rather than being dominated by the larger NC subset.

The final model architecture and pooling parameters were selected through controlled tests of the graph-level pooling representation, hidden dimension, and parameters governing the distance-weighted local pooling. The robustness of the resulting model with respect to stochastic parameter initialization and mini-batch ordering was subsequently evaluated using five independent training seeds while retaining the same fixed training/validation partition. The corresponding results are reported below.

\subsection{Model selection and robustness}
\subsubsection{Pooling strategy}

We first assessed how the construction of the graph-level representation affects the classification of exact adsorption-site retention. Three pooling strategies were compared: (i) global mean pooling over the complete local Ni subgraph, $\mathbf{g}_{\mathrm{global}}$; (ii) separate O- and H-centered local representations, $\mathbf{g}_{\mathrm{O}}\Vert\mathbf{g}_{\mathrm{H}}$; and (iii) a mixed representation combining the global and local embeddings, $\mathbf{g}_{\mathrm{global}}\Vert\mathbf{g}_{\mathrm{O}}\Vert\mathbf{g}_{\mathrm{H}}$. The definitions of the corresponding pooling operations are given in Section \ref{sec:si_architecture}. All three variants were trained using the same fixed training/validation partition, random seed, message-passing architecture, and optimization settings, such that only the graph-level pooling representation was varied.

The three pooling strategies showed comparable overall discrimination performance but different behavior at the operational threshold $s_{\mathrm{ret}}=0.80$ (Table \ref{tab:retention_pooling}). This threshold was chosen as a compromise between precision and recall that deliberately favors precision: in screening, configurations predicted to be retained are intended for subsequent DFT relaxation, so limiting the number of migrating candidates that pass the filter avoids wasted calculations, at the cost of rejecting a fraction of configurations that would in fact retain their binding sites. The mixed global--local representation yielded the lowest domain-macro validation BCE (0.135) and the highest AP (0.943) and ROC--AUC (0.982), while the global representation achieved the highest recall (0.816) at the lowest precision (0.903). The local O/H-centered representation provided the highest precision (0.938), with a recall of 0.733, although its margin in precision over the mixed representation (0.930) was small. In selecting the representation, we sought a balance between precision and recall: the classifier should limit the number of migrating candidates passing the filter without rejecting an excessive fraction of motif-retaining configurations. Both the local and the mixed representations satisfy this criterion, whereas the global representation gains recall at the expense of precision. Either the local or the mixed representation would therefore be a suitable choice; the local representation, which yielded the highest precision, was selected for the subsequent architecture tests.

\begin{table}[H]
\centering
\caption{Effect of the pooling strategy on the motif-retention model evaluated on the fixed validation set. Precision and recall are reported at $s_{\mathrm{ret}}=0.80$.}
\label{tab:retention_pooling}
\renewcommand{\arraystretch}{1.2}
\setlength{\tabcolsep}{10pt}
\begin{tabular}{|c|c|c|c|c|c|}
\hline
Pooling & \makecell{Domain-macro\\BCE} & AP & ROC--AUC & Precision & Recall \\
\hline
$\mathbf{g}_{\mathrm{global}}$ & 0.175& 0.926& 0.968& 0.903& 0.816\\
\hline
$\mathbf{g}_{\mathrm{O}}\Vert\mathbf{g}_{\mathrm{H}}$ & 0.143& 0.929& 0.978& 0.938& 0.733\\
\hline
$\mathbf{g}_{\mathrm{global}}\Vert\mathbf{g}_{\mathrm{O}}\Vert\mathbf{g}_{\mathrm{H}}$ & 0.135& 0.943& 0.982& 0.930& 0.777\\
\hline
\end{tabular}
\end{table}

We further examined the two parameters controlling the local pooling operation: the spatial decay length $\sigma_p$ and the explicit anchor boost $\beta$ (see Eqs. \ref{eq:pool_O}--\ref{eq:pool_H}). A grid search was performed using \(\sigma_p \in \{1.0, 1.5, 2.0\}\) \AA{} and \(\beta \in \{0, 0.5, 1.0, 2.0\}\), and $\sigma_p = 1.0$ \AA{} and $\beta = 1.0$ were retained for the final local-pooling representation.

\subsubsection{Model capacity}

The effect of model capacity was assessed by varying the hidden dimension $d$ while keeping the local-pooling parameters, training/validation partition, and random seed fixed. The $d=64$ architecture achieved the lowest domain-macro validation BCE (0.143) and the highest AP (0.929), with a ROC--AUC (0.978) equal to that of $d=256$ (Table \ref{tab:retention_capacity}). At the operational threshold $s_{\mathrm{ret}}=0.80$, $d=64$ also yielded the highest recall (0.733), with a precision of 0.938. Increasing the hidden dimension to $d=128$ gave a marginally higher precision (0.941) but a substantially lower recall (0.617), while $d=256$ offered no improvement in either quantity (precision 0.926, recall 0.665). Since $d=64$ provided the best balance between precision and recall, it was selected for the final architecture.

\begin{table}[H]
\centering
\caption{Effect of the hidden dimension on the motif-retention model evaluated on the fixed validation set. Precision and recall are reported at $s_{\mathrm{ret}}=0.80$.}
\label{tab:retention_capacity}
\renewcommand{\arraystretch}{1.2}
\setlength{\tabcolsep}{10pt}
\begin{tabular}{|c|c|c|c|c|c|c|}
\hline
$d$ & Parameters & \makecell{Domain-macro\\BCE} & AP & ROC--AUC & Precision & Recall \\
\hline
64  & 84,229& 0.143& 0.929& 0.978& 0.938& 0.733\\
\hline
128 & 291,077& 0.169& 0.895& 0.964& 0.941& 0.617\\
\hline
256 & 1,073,413 & 0.182& 0.921& 0.978& 0.926& 0.665\\
\hline
\end{tabular}
\end{table}

\subsubsection{Random-seed sensitivity}

The robustness of the final architecture with respect to stochastic training effects was assessed using five independent training seeds while keeping the training/validation partition and all hyperparameters fixed. Performance remained relatively consistent across runs, with a domain-macro validation BCE of $0.151\pm0.017$, an AP of $0.914\pm0.023$, and a ROC--AUC of $0.970\pm0.009$ (mean $\pm$ SD; Table \ref{tab:retention_seeds}). At the operational threshold $s_{\mathrm{ret}}=0.80$, precision remained high ($0.953\pm0.018$), while recall showed greater variability ($0.710\pm0.062$). Seed 10, adopted as the reference model, yielded a precision of 0.949, a recall of 0.718, an AP of 0.924, and a ROC--AUC of 0.968, all within one standard deviation of the corresponding means, whereas its domain-macro validation BCE (0.171) was the highest among the five seeds. Seed 42, which achieved the lowest domain-macro validation BCE (0.131) while balancing a precision of 0.950 with a recall of 0.733, would represent an equally suitable choice.

\begin{table}[H]
\centering
\caption{Sensitivity of the motif-retention model to the training seed on the fixed validation set. Precision and recall are reported at $s_{\mathrm{ret}}=0.80$.}
\label{tab:retention_seeds}
\renewcommand{\arraystretch}{1.2}
\setlength{\tabcolsep}{2.8pt}
\begin{tabular}{|c|c|c|c|c|c|}
\hline
Seed & \makecell{Domain-macro\\BCE} & AP & ROC--AUC & Precision & Recall \\
\hline
2  & 0.142 & 0.916 & 0.970 & 0.946 & 0.762\\
\hline
10 & 0.171 & 0.924 & 0.968 & 0.949 & 0.718\\
\hline
22 & 0.166 & 0.874 & 0.957 & 0.984 & 0.602\\
\hline
32 & 0.143 & 0.929 & 0.978 & 0.938 & 0.733 \\
\hline
42 & 0.131 & 0.927 & 0.979 & 0.950 & 0.733 \\
\hline\hline
\textbf{Mean $\pm$ SD}
& $0.151\pm0.017$
& $0.914\pm0.023$
& $0.970\pm0.009$
& $0.953\pm0.018$
& $0.710\pm0.062$ \\
\hline
\end{tabular}
\end{table}

\subsection{Model performance and transfer to MACE-relaxed inputs}
\label{sec:si_class_perf}

The selected motif-retention architecture showed a clear separation between retained and migrated configurations when evaluated using DFT-derived structural inputs. Over the complete validation set, the DFT-input model yielded an AP of 0.924 and a ROC--AUC of 0.968, with similarly strong performance for the nanocluster and surface subsets (Table \ref{tab:ret_performance_dft_vs_ml}). At the operational threshold $s_{\mathrm{ret}}=0.80$, the model achieved a precision of 0.949, a recall of 0.718, and a specificity of 0.992.

To make the retention model consistent with the scalable workflow adopted for energy prediction, the same architecture was subsequently retrained using MACE-relaxed Ni geometries as structural inputs, while retaining the same DFT-defined retention labels and fixed training/validation partition. Replacing the DFT-derived inputs with their MACE-relaxed counterparts largely preserved the overall discrimination performance, with an AP of 0.911 and a ROC--AUC of 0.972 over the complete validation set. At $s_{\mathrm{ret}}=0.80$, precision decreased slightly to 0.933 and recall to 0.680, while specificity remained high (0.990). Similar trends were observed for the nanocluster subset, for which the ROC--AUC increased slightly (0.971 versus 0.966), whereas the surface subset showed moderate reductions in AP, ROC--AUC, and recall, with nearly unchanged precision and unchanged specificity. For the complete validation set, these differences are smaller than the seed-to-seed standard deviations obtained with DFT-derived inputs (Table \ref{tab:retention_seeds}).

Overall, these results indicate that MACE-relaxed structural inputs largely preserve the classifier's ability to identify motif-retaining configurations, with performance comparable to DFT-derived inputs. Importantly, this approach eliminates the need for DFT relaxation of the bare catalyst at inference, enabling efficient screening of larger systems. The MACE-input retention model was therefore adopted together with M6-MACE for the subsequent screening and structural analyses.

\begin{table}[H]
\centering
\caption{Motif-retention performance using DFT- and MACE-relaxed structural inputs on the same fixed validation set, with identical labels, seed, and hyperparameters. Precision, recall, and specificity are reported at $s_{\mathrm{ret}}=0.80$.}
\label{tab:ret_performance_dft_vs_ml}
\renewcommand{\arraystretch}{1.2}
\setlength{\tabcolsep}{5.5pt}
\begin{tabular}{|c|c|c|c|c|c|c|c|}
\hline
Domain & $N_{\mathrm{val}}$ & Geometry & AP & ROC--AUC & Precision & Recall & Specificity \\
\hline
\multirow{2}{*}{All}           & \multirow{2}{*}{1219}& DFT & 0.924& 0.968& 0.949& 0.718& 0.992\\
                               &                       & MACE  & 0.911& 0.972& 0.933& 0.680& 0.990\\
\hline
\multirow{2}{*}{Nanoclusters} & \multirow{2}{*}{1154}& DFT & 0.918& 0.966& 0.949& 0.710& 0.993\\
                               &                       & MACE  & 0.905& 0.971& 0.932& 0.672& 0.991\\
\hline
\multirow{2}{*}{Surfaces}      & \multirow{2}{*}{65}& DFT & 0.973& 0.983& 0.947& 0.783& 0.976\\
                               &                       & MACE  & 0.963& 0.973& 0.944& 0.739& 0.976\\
\hline
\end{tabular}
\end{table}

\section[System-resolved performance of the final M6-MACE model]{System-resolved performance of the final\\ M6-MACE model}
\label{sec:M6_energy}

\begin{figure}[H]
    \centering
    \includegraphics[width=1.0\linewidth]{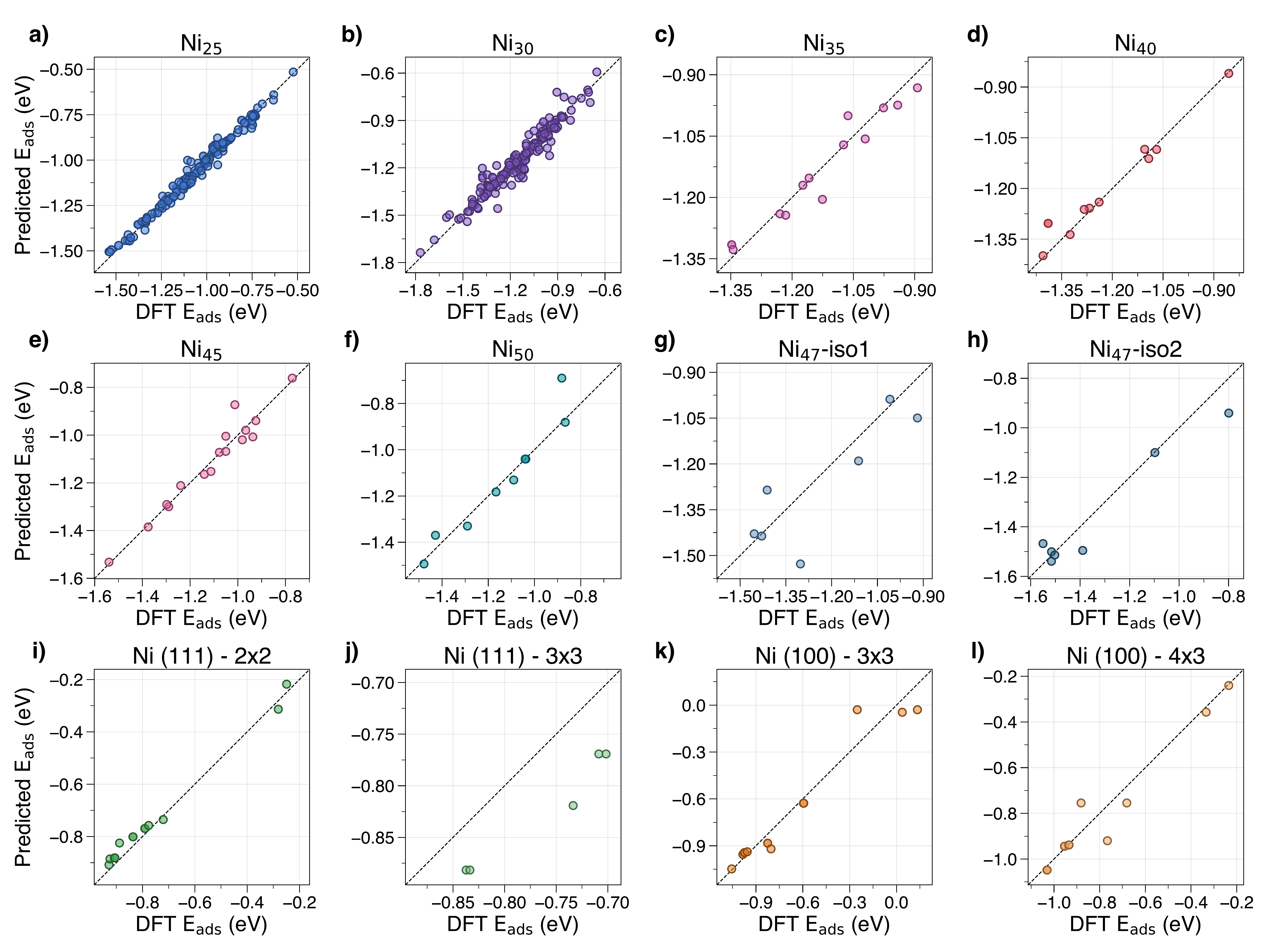}
    \caption{Comparison between DFT-reference and GNN-predicted absolute adsorption energies, $\mathrm{E}_{\mathrm{ads}}$, obtained with the final M6-MACE model for the validation configurations of (a--h) Ni nanoclusters and (i--l) periodic Ni surfaces. Panels (a--h) correspond to Ni$_{25}$, Ni$_{30}$, Ni$_{35}$, Ni$_{40}$, Ni$_{45}$, Ni$_{50}$, Ni$_{47}$-iso1, and Ni$_{47}$-iso2, respectively. Panels (i,j) correspond to Ni(111)-2$\times$2 and Ni(111)-3$\times$3, while panels (k,l) correspond to Ni(100)-3$\times$3 and Ni(100)-4$\times$3. The dashed diagonal line indicates perfect agreement between DFT and predicted values.}
    \label{fig:M6_abs_en}
\end{figure}

\begin{figure}[H]
    \centering
    \includegraphics[width=1.0\linewidth]{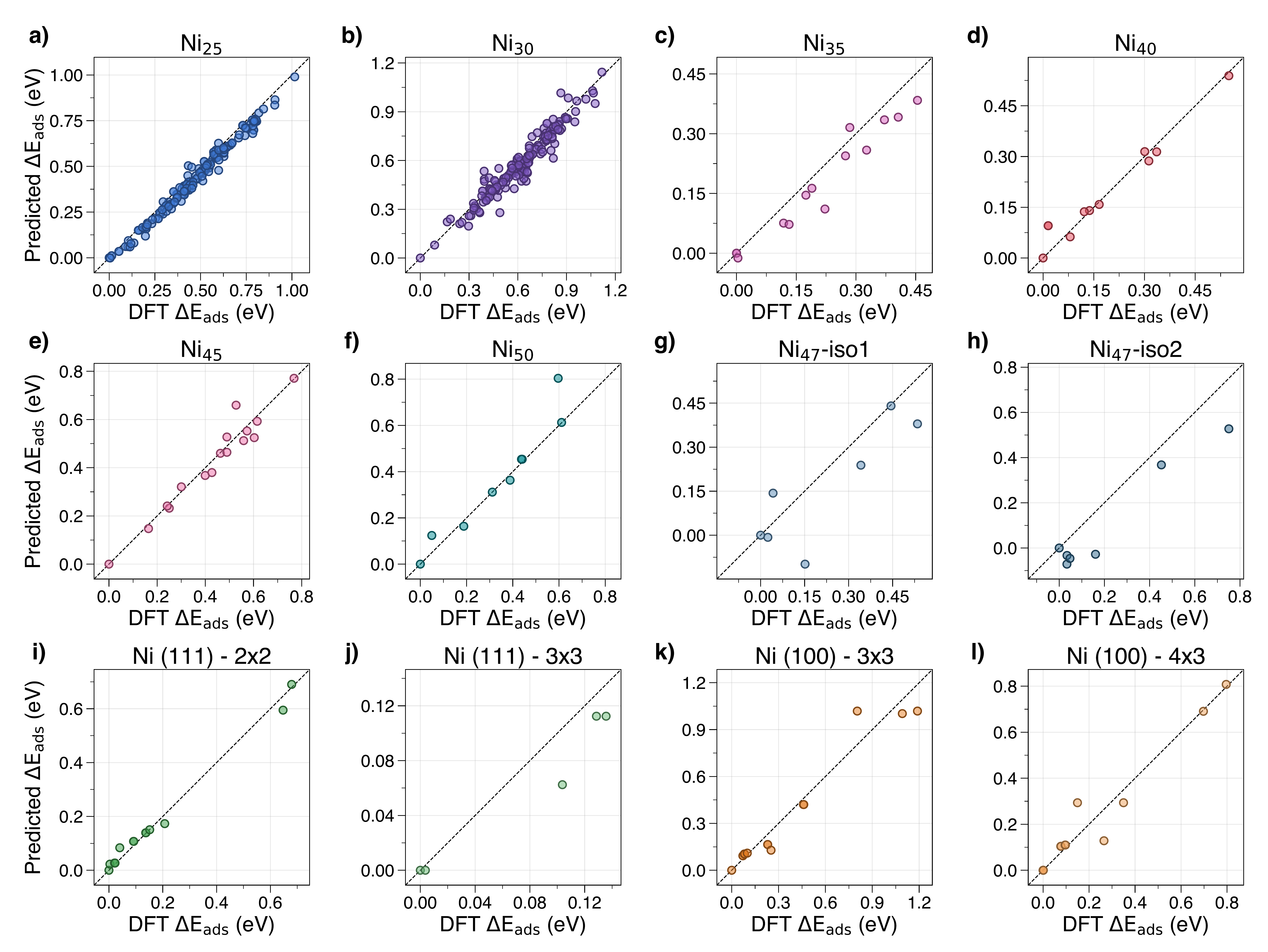}
    \caption{Comparison between DFT-reference and GNN-predicted relative adsorption energies, $\Delta \mathrm{E}_{\mathrm{ads}}$, obtained with the final M6-MACE model for the validation configurations of (a--h) Ni nanoclusters and (i--l) periodic Ni surfaces. Relative energies are referenced to the lowest-energy DFT configuration within each system, following Eqs. 3--4 of the main text. Panels (a--h) correspond to Ni$_{25}$, Ni$_{30}$, Ni$_{35}$, Ni$_{40}$, Ni$_{45}$, Ni$_{50}$, Ni$_{47}$-iso1, and Ni$_{47}$-iso2, respectively. Panels (i,j) correspond to Ni(111)-2$\times$2 and Ni(111)-3$\times$3, while panels (k,l) correspond to Ni(100)-3$\times$3 and Ni(100)-4$\times$3. The dashed diagonal line indicates perfect agreement between DFT and predicted values.}
    \label{fig:M6_rel_en}
\end{figure}

\begin{table}[H]
\centering
\caption{Performance of the final M6-MACE energy model for the individual nanocluster and periodic-surface systems in the validation set. Mean absolute errors (MAEs) and root-mean-square errors (RMSEs) are reported for both absolute adsorption energies, $\mathrm{E}_{\mathrm{ads}}$, and relative adsorption energies, $\Delta \mathrm{E}_{\mathrm{ads}}$. $N_{\mathrm{val}}$ denotes the number of validation configurations for each system.}
\label{tab:M6_system_performance}
\renewcommand{\arraystretch}{1.35}
\setlength{\tabcolsep}{12pt}
\begin{tabular}{|l|c|l|c|c|}
\hline
\textbf{System} & \textbf{$N_{\mathrm{val}}$} & \textbf{Metric} & \textbf{$\mathrm{E}_{\mathrm{ads}}$ (eV)} & \textbf{$\Delta \mathrm{E}_{\mathrm{ads}}$ (eV)} \\
\hline
\multirow{2}{*}{Ni$_{25}$} & \multirow{2}{*}{161} & MAE & 0.019 & 0.039 \\ & & RMSE & 0.026 & 0.045 \\ \hline
\multirow{2}{*}{Ni$_{30}$} & \multirow{2}{*}{150} & MAE & 0.039 & 0.053 \\ & & RMSE & 0.056 & 0.067 \\ \hline
\multirow{2}{*}{Ni$_{35}$} & \multirow{2}{*}{14} & MAE & 0.027 & 0.042 \\ & & RMSE & 0.035 & 0.051 \\ \hline
\multirow{2}{*}{Ni$_{40}$} & \multirow{2}{*}{11} & MAE & 0.026 & 0.025 \\ & & RMSE & 0.039 & 0.037 \\ \hline
\multirow{2}{*}{Ni$_{45}$} & \multirow{2}{*}{16} & MAE & 0.030 & 0.032 \\ & & RMSE & 0.045 & 0.045 \\ \hline
\multirow{2}{*}{Ni$_{50}$} & \multirow{2}{*}{9} & MAE & 0.042 & 0.040 \\ & & RMSE & 0.070 & 0.075 \\ \hline
\multirow{2}{*}{Ni$_{47}$-iso1} & \multirow{2}{*}{7} & MAE & 0.088 & 0.092 \\ & & RMSE & 0.114 & 0.125 \\ \hline
\multirow{2}{*}{Ni$_{47}$-iso2} & \multirow{2}{*}{7} & MAE & 0.054 & 0.109 \\ & & RMSE & 0.075 & 0.129 \\ \hline\hline
\multirow{2}{*}{Ni(111)-2$\times$2} & \multirow{2}{*}{14} & MAE & 0.029 & 0.015 \\ & & RMSE & 0.031 & 0.022 \\ \hline
\multirow{2}{*}{Ni(111)-3$\times$3} & \multirow{2}{*}{5} & MAE & 0.061 & 0.017 \\ & & RMSE & 0.063 & 0.022 \\ \hline
\multirow{2}{*}{Ni(100)-3$\times$3} & \multirow{2}{*}{15} & MAE & 0.062 & 0.062 \\ & & RMSE & 0.086 & 0.087 \\ \hline
\multirow{2}{*}{Ni(100)-4$\times$3} & \multirow{2}{*}{13} & MAE & 0.037 & 0.034 \\ & & RMSE & 0.060 & 0.059 \\ \hline
\end{tabular}
\end{table}

\section{Comparison with descriptor-based regression models}
\label{sec:si_baselines}

To assess whether the local graph representation provides predictive information beyond conventional adsorption-site descriptors, we compared M6-MACE with two simpler regression models: Ridge regression\cite{Hoerl_1970} and XGBoost\cite{Chen_2016}. Both baselines were trained to reproduce the same DFT adsorption energies using four descriptors derived exclusively from the bare Ni geometry: the co-adsorption motif, the generalized coordination numbers of the O- and H-binding sites ($\mathrm{GCN_O}$ and $\mathrm{GCN_H}$), and the distance between the corresponding adsorption-site centroids ($d_{\mathrm{OH}}$).

The generalized coordination number (GCN) describes the local coordination environment of an adsorption site by accounting for the coordination numbers of its neighboring metal atoms\cite{Calle-Vallejo_2014,Calle-Vallejo_2015}. For a site defined by a set $S$ of Ni anchor atoms, it is calculated as
\begin{equation}
\mathrm{GCN}(S) = \frac{1}{CN_{\max}(S)} \sum_{j \in \mathrm{NN}(S)} CN(j),
\label{eq:gcn}
\end{equation}
where $\mathrm{NN}(S)$ is the union of the first-shell Ni neighbors of the anchor atoms, excluding the anchors themselves and counting shared neighbors only once. Here, $CN(j)$ is the conventional coordination number of neighboring atom $j$, while $CN_{\max}(S)$ is the site-dependent normalization factor, equal to 12, 18, 22, and 26 for top, bridge, threefold hollow, and fourfold sites, respectively. The descriptors $\mathrm{GCN_O}$ and $\mathrm{GCN_H}$ correspond to the O- and H-binding sites identified from their respective Ni anchor sets. The site--site distance $d_{\mathrm{OH}}$ is calculated between the corresponding anchor centroids, accounting for periodic boundary conditions when applicable. All descriptors were evaluated on the MACE-relaxed bare Ni structures, without using explicit adsorbate coordinates.

Unlike M6-MACE--obtained by fine-tuning the M5 checkpoint--the descriptor-based baselines were trained from scratch using the same M6 training split. The baseline training set comprised 1,659 configurations: 1,490 NC configurations and 169 periodic-surface configurations. Hyperparameter selection and evaluation were performed using the original M6 validation set, comprising 422 configurations: 375 nanoclusters and 47 periodic surfaces. As in M6-MACE training, NC and periodic-surface configurations were assigned weights of 1 and 4, respectively, while hyperparameter selection was based on the unweighted validation MAE. Because the descriptor-based baselines were trained from scratch rather than initialized from the preceding M5 model, the comparison evaluates the predictive performance of the complete modeling approaches rather than isolating the effect of the graph representation alone.

Ridge regression was used as a linear, physically motivated reference model. Its feature representation included the standardized continuous descriptors $\mathrm{GCN_O}$, $\mathrm{GCN_H}$, and $d_{\mathrm{OH}}$, an interaction term $\mathrm{GCN_O}\times\mathrm{GCN_H}$, one-hot-encoded adsorption motifs, and motif-dependent $\mathrm{GCN_O}$ terms. The latter allow different adsorption motifs to have distinct intercepts and coordination sensitivities. The continuous descriptors were standardized using the means and standard deviations calculated from the training dataset. The regularization parameter was selected from $\alpha \in \{0.01,0.1,1,10,100\}$.

XGBoost was employed as a nonlinear regression baseline, allowing more complex dependencies and interactions between the same structural descriptors to be learned directly. The input representation comprised the three continuous descriptors and the one-hot-encoded adsorption motifs, without explicitly introducing interaction terms. The hyperparameter search considered maximum tree depths of 2, 3, and 4, learning rates of 0.03 and 0.05, and L2 regularization parameters of 1, 2, and 5. All models used 500 estimators, a sub-sampling fraction of 0.9, and a column-sampling fraction of 1.0. For both baselines, the random seed was fixed to 32, and hyperparameters were selected by minimizing the unweighted validation MAE.

\subsection{Internal validation}

\begin{figure}[H]
    \centering
    \includegraphics[width=1.0\linewidth]{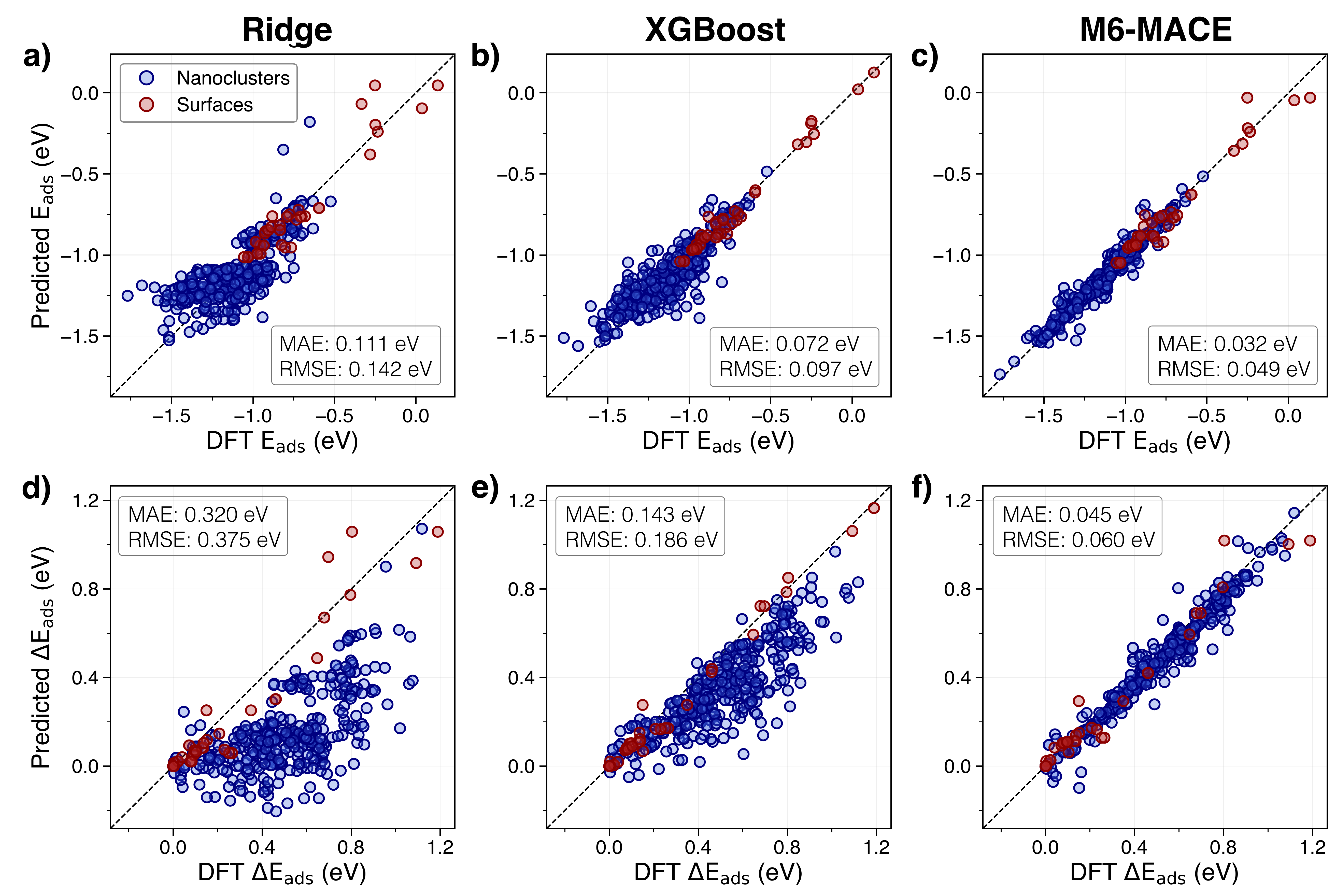}
    \caption{Comparison of descriptor-based regression models with M6-MACE on the internal validation set. The top row shows absolute adsorption energies, $\mathrm{E}_{\mathrm{ads}}$, predicted by (a) Ridge, (b) XGBoost, and (c) M6-MACE; the bottom row shows the corresponding relative adsorption energies, $\Delta\mathrm{E}_{\mathrm{ads}}$, for (d) Ridge, (e) XGBoost, and (f) M6-MACE. Nanocluster and periodic-surface configurations are shown in blue and red, respectively. Relative energies are referenced to the lowest-energy DFT configuration within each system, following Eqs. 3--4 of the main text. The mean absolute error (MAE) and root-mean-square error (RMSE) are reported in each panel, and the dashed diagonal lines indicate perfect agreement.}
    \label{fig:si_baseline_validation}
\end{figure}

Figure \ref{fig:si_baseline_validation} compares Ridge, XGBoost, and M6-MACE on the internal validation set. For absolute adsorption energies, M6-MACE achieves an MAE of 0.032 eV and an RMSE of 0.049 eV, substantially lower than Ridge (0.111 and 0.142 eV) and XGBoost (0.072 and 0.097 eV). The improvement is particularly pronounced for NC configurations, where the MAEs are 0.117, 0.077, and 0.031 eV and the RMSEs 0.147, 0.102, and 0.047 eV for Ridge, XGBoost, and M6-MACE, respectively. For periodic surfaces, XGBoost outperforms M6-MACE (MAE of 0.030 versus 0.045 eV; RMSE of 0.042 versus 0.064 eV), whereas Ridge yields 0.064 and 0.090 eV.

The difference becomes more pronounced when considering relative adsorption energies, $\Delta\mathrm{E}_{\mathrm{ads}}$, referenced within each system to its lowest-energy DFT configuration. M6-MACE achieves an MAE of 0.045 eV and an RMSE of 0.060 eV, compared with 0.320 and 0.375 eV for Ridge and 0.143 and 0.186 eV for XGBoost. The descriptor-based models reproduce part of the overall energetic trends but show substantially larger deviations in the energy differences between configurations, particularly for NCs, where the MAEs are 0.352, 0.158, and 0.046 eV and the RMSEs 0.396, 0.197, and 0.060 eV for Ridge, XGBoost, and M6-MACE, respectively. For periodic surfaces, XGBoost again yields lower relative-energy errors (MAE of 0.025 eV; RMSE of 0.038 eV) than M6-MACE (0.035 and 0.060 eV), whereas Ridge yields 0.065 and 0.096 eV. These results show that M6-MACE resolves configurational energy differences that are not fully reproduced by models based on the adsorption motif, generalized coordination numbers, and site--site distance alone.

\subsection{Transferability to external test systems}

\begin{figure}[H]
    \centering
    \includegraphics[width=1.0\linewidth]{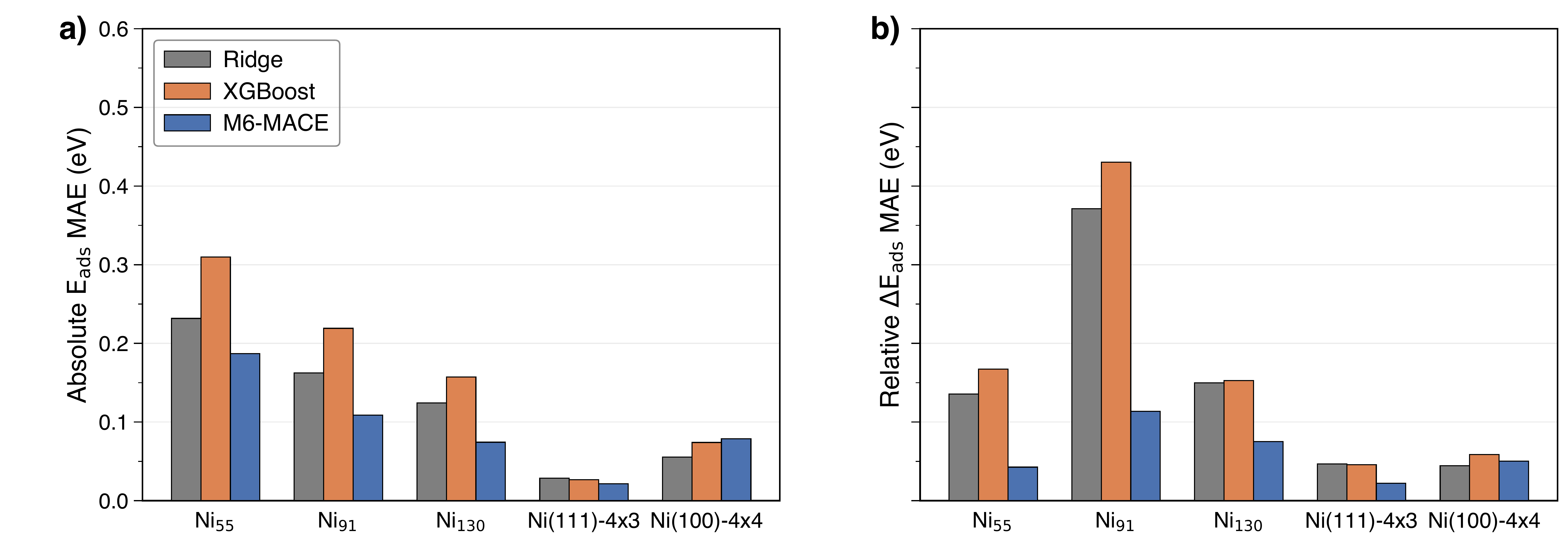}
    \caption{Comparison of the predictive accuracy of Ridge, XGBoost, and M6-MACE on five external test systems: Ni$_{55}$, Ni$_{91}$, Ni$_{130}$, Ni(111)-4$\times$3, and Ni(100)-4$\times$4. Mean absolute errors (MAEs) are reported for (a) absolute adsorption energies, $\mathrm{E}_{\mathrm{ads}}$, and (b) relative adsorption energies, $\Delta\mathrm{E}_{\mathrm{ads}}$. All models are evaluated on the same selected DFT-reference configurations, with relative energies referenced to the lowest-energy DFT configuration within each system.}
    \label{fig:si_baseline_external}
\end{figure}

The descriptor-based baselines were further evaluated on the same five external test systems used to assess M6-MACE: Ni$_{55}$, Ni$_{91}$, Ni$_{130}$, Ni(111)-4$\times$3, and Ni(100)-4$\times$4. These systems were excluded from training and hyperparameter selection. All three models were evaluated on the same 199 DFT-relaxed configurations, using identical final adsorption motifs and O- and H-anchor sets. The corresponding descriptors were calculated on the MACE-relaxed bare Ni geometries, following the same evaluation protocol adopted for M6-MACE. Figure \ref{fig:si_baseline_external} compares the absolute- and relative-energy MAEs across the five systems.

For absolute adsorption energies, M6-MACE generally outperforms both descriptor-based baselines, with the largest improvements observed for NCs. For relative adsorption energies, M6-MACE achieves the lowest MAE for Ni$_{55}$, Ni$_{91}$, Ni$_{130}$, and Ni(111)-4$\times$3, whereas on Ni(100)-4$\times$4 Ridge performs slightly better (0.045 versus 0.050 eV). The advantage is particularly pronounced for Ni$_{55}$, Ni$_{91}$, and Ni(111)-4$\times$3, whereas the three models perform more similarly on Ni(100)-4$\times$4. The external comparison therefore confirms that M6-MACE is particularly effective at reproducing adsorption-energy differences across structurally diverse configurations, including previously unseen systems.

Overall, the baseline comparison demonstrates that conventional coordination-based descriptors capture important trends in dissociative adsorption energetics but do not fully reproduce the configurational energy differences resolved by M6-MACE. The improvement is most pronounced for structurally diverse NC environments, whereas simpler descriptors can provide competitive predictions for some of the more regular periodic surfaces. Because M6-MACE and the baselines employ different training strategies, this comparison evaluates the performance of the complete models rather than isolating the effect of the graph architecture alone.

\section{Transferability to larger nanoclusters and surfaces}

\begin{figure}[H]
    \centering
    \includegraphics[width=1.0\linewidth]{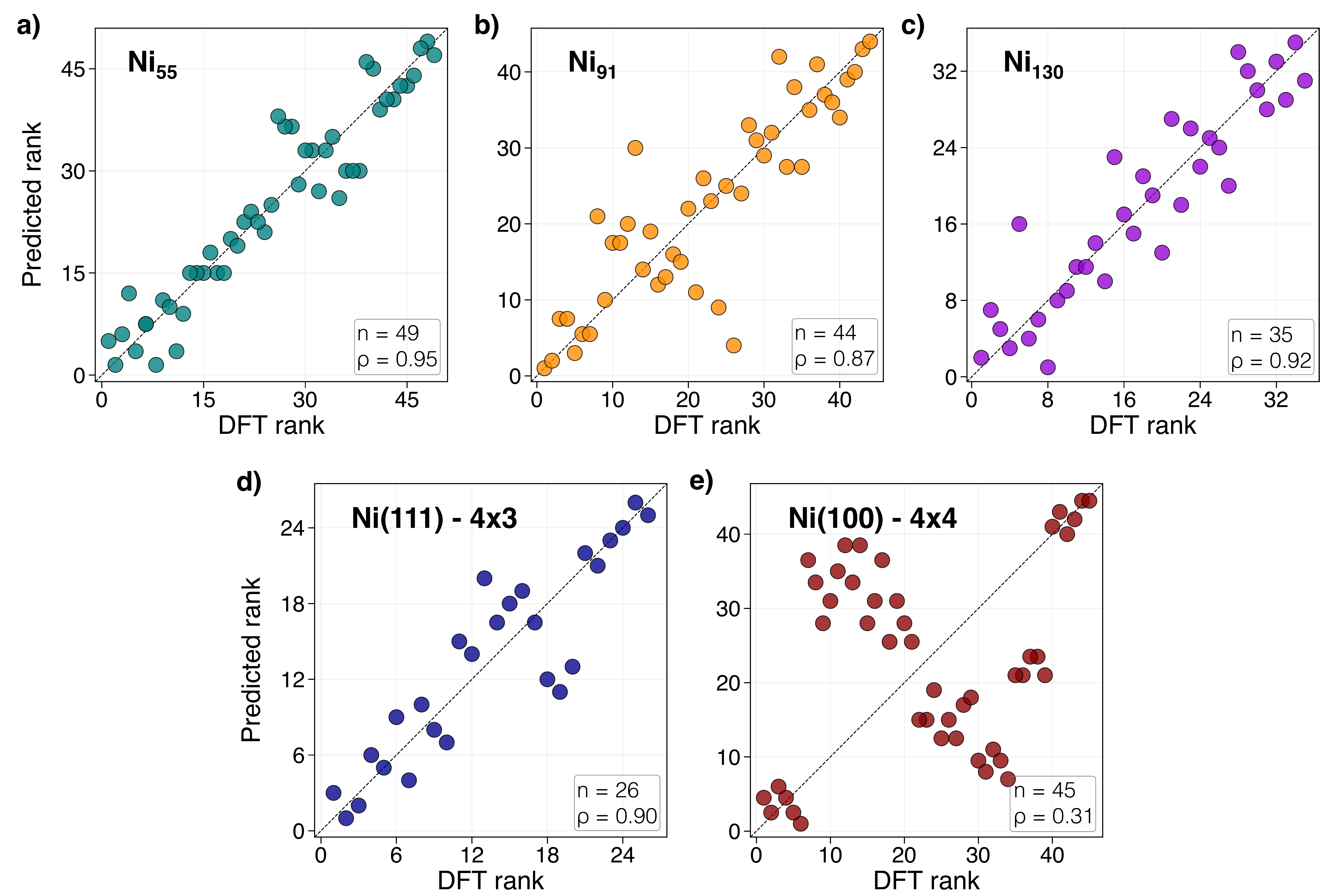}
    \caption{Comparison between DFT-reference and GNN-predicted ranking orders for selected configurations of three external test nanoclusters: (a) Ni$_{55}$, (b) Ni$_{91}$, and (c) Ni$_{130}$; and two extended periodic surfaces: (d) Ni(111)-4$\times$3 and (e) Ni(100)-4$\times$4. The Spearman rank correlation coefficient ($\rho$) is reported in each panel's legend, and the dashed diagonal line indicates perfect agreement.}
    \label{fig:transferability_rank}
\end{figure}

\begin{table}[H]
\centering
\caption{Performance of the retention classifier on the independent test systems. Accuracy is the fraction of configurations correctly classified as retained or migrated. Retained-class precision is the fraction of configurations predicted to be retained that preserve both their adsorption motif and the O/H anchoring sites after DFT relaxation.}
\label{tab:retention_external}
\renewcommand{\arraystretch}{1.3}
\setlength{\tabcolsep}{15pt}
\begin{tabular}{|l|r|r|}
\hline
System & Accuracy (\%) & Retained-class precision (\%) \\
\hline
Ni$_{55}$             & 89.8 (44/49) & 96.0 (24/25) \\ \hline
Ni$_{91}$             & 84.1 (37/44) & 96.0 (24/25) \\ \hline
Ni$_{130}$            & 77.1 (27/35) & 82.6 (19/23)\\ \hline
Ni(111)-4$\times$3    & 92.3 (24/26) & 91.7 (11/12) \\ \hline
Ni(100)-4$\times$4    & 93.3 (42/45) & 97.5 (39/40) \\ \hline
\end{tabular}
\end{table}

\label{sec:si_screening_cost}
\begin{table}[H]
\centering
\caption{Computational cost of screening OH$^*$+H$^*$ co-adsorption configurations on representative finite Ni nanosystems of increasing size. GNN evaluations were performed on a single GPU, while candidate-site enumeration, geometric filtering, and pre-processing were performed on the CPU. $N_{\mathrm{conf}}$ denotes the number of candidate O/H site pairs evaluated by the GNNs. The reported wall-clock times exclude the preceding MACE geometry optimizations, which were on the order of minutes for all systems, and include adsorption-site enumeration and filtering, local graph construction, adsorption-energy prediction, motif-retention classification, and ranking.}
\label{tab:screening_times}
\renewcommand{\arraystretch}{1.3}
\setlength{\tabcolsep}{15pt}
\begin{tabular}{|l|r|r|}
\hline
\textbf{Ni system} & \textbf{$N_{\mathrm{conf}}$} & \textbf{Wall time (min)} \\
\hline
Ni$_{55}$   & 19,680  & 4.6   \\ \hline
Ni$_{91}$   & 26,940  & 8.5   \\ \hline
Ni$_{130}$  & 46,378  & 18.0  \\ \hline
Ni$_{442}$  & 87,178  & 54.0  \\ \hline
Ni$_{1141}$ & 181,876 & 160.4 \\ \hline
Ni$_{3007}$ & 146,914 & 254.8 \\ \hline
\end{tabular}
\end{table}

\section{Structural determinants of adsorption-site stability}
\label{sec:si_structural}

Throughout this section, retained configurations are those assigned a motif-retention score $s_{\mathrm{ret}} \ge 0.80$ by the classifier (Section \ref{sec:si_retention}).

\subsection{Structural descriptors}
\label{sec:si_descriptors}

The adsorption-site generalized coordination number (GCN) used to characterize the O- and H-binding environments is defined in Eq. \ref{eq:gcn} (Section \ref{sec:si_baselines}). For comparisons of surface coordination across systems, an atomic GCN is calculated separately for each surface Ni atom, using $CN_{\max}=12$. The mean surface GCN is obtained by averaging these atomic values over all surface Ni atoms.

Local crystalline order is quantified by the averaged Steinhardt bond-orientational order parameter $\bar{q}_6$\cite{Steinhardt_1983,Lechner_2008}. For each surface atom $i$,
\begin{equation}
q_{lm}(i) = \frac{1}{N_b(i)} \sum_{j \in \mathcal{N}(i)} Y_{lm}(\hat{\mathbf{r}}_{ij}),
\label{eq:qlm}
\end{equation}
\begin{equation}
\bar{q}_{lm}(i) = \frac{1}{N_b(i)+1}\Bigl( q_{lm}(i) + \sum_{j \in \mathcal{N}(i)} q_{lm}(j) \Bigr),
\label{eq:qlm_avg}
\end{equation}
\begin{equation}
\bar{q}_l(i) = \sqrt{\frac{4\pi}{2l+1} \sum_{m=-l}^{l} \bigl|\bar{q}_{lm}(i)\bigr|^2},
\label{eq:qbar}
\end{equation}
with $l=6$, where $Y_{lm}$ are the spherical harmonics, $\hat{\mathbf{r}}_{ij}$ is the unit vector from $i$ to neighbor $j$, $\mathcal{N}(i)$ denotes its first-shell Ni neighbors, and $N_b(i)$ their number. The mean surface $\bar{q}_6$ is obtained by averaging $\bar{q}_6(i)$ over all surface Ni atoms.

\subsection{O- and H-site dependence for individual nanoclusters}
\label{sec:si_oh_dependence}

\begin{figure}[H]
    \centering
    \includegraphics[width=1.0\linewidth]{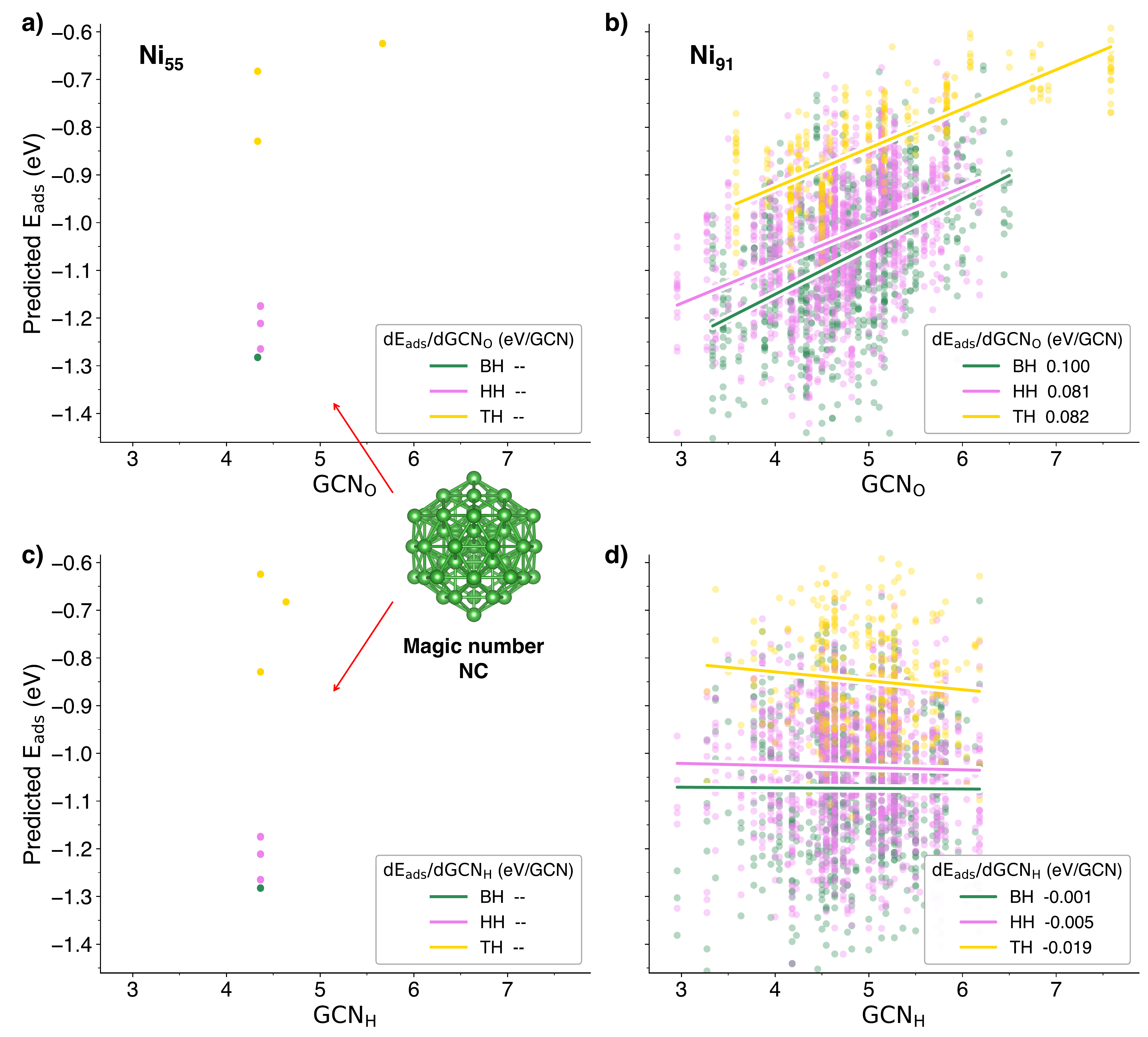}
    \caption{Dependence of the predicted adsorption energy on the O- and H-site generalized coordination numbers (GCNs) for Ni$_{55}$ (a,c) and Ni$_{91}$ (b,d), resolved by co-adsorption motif. (a,b) $\mathrm{E}_{\mathrm{ads}}$ versus $\mathrm{GCN_O}$ and (c,d) versus $\mathrm{GCN_H}$, with per-motif slopes (eV per GCN unit) in the legends. Ni$_{55}$ is a high-symmetry magic-number icosahedron (inset) exposing only a few, nearly degenerate coordination environments, so motif-resolved slopes cannot be meaningfully fitted.}
    \label{fig:si_gcn_ni5591}
\end{figure}

\subsection{Disentangling the O and H contributions}
\label{sec:si_disentangle}

\begin{figure}[H]
    \centering
    \includegraphics[width=1.0\linewidth]{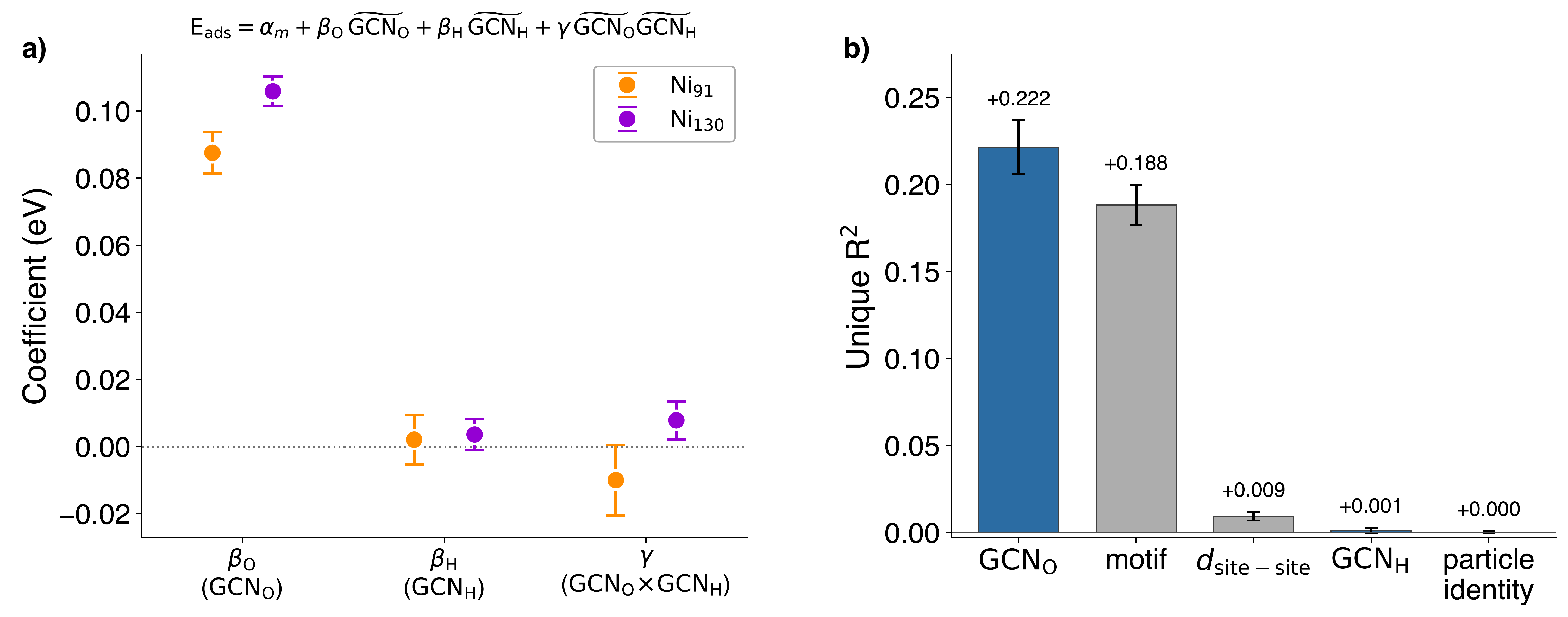}
    \caption{Statistical analysis of the contributions of the O- and H-site environments to the predicted adsorption energy. (a) Coefficients of the multivariable fit of Eq. \ref{eq:multivar} for Ni$_{91}$ (orange) and Ni$_{130}$ (purple). (b) Unique contributions to the coefficient of determination, $\Delta \mathrm{R}^2$ (Eq. \ref{eq:unique_r2}), for each predictor block in the pooled Ni$_{91}$ and Ni$_{130}$ configurations. Error bars in (b) indicate the standard deviation across 15 train/test splits.}
    \label{fig:si_residual}
\end{figure}

To disentangle the contributions of the O- and H-binding environments, two complementary analyses were performed on the predicted adsorption energies of Ni$_{91}$ and Ni$_{130}$ (Figure \ref{fig:si_residual}). First, the adsorption energies of each NC were fitted independently to the multivariable model
\begin{equation}
\mathrm{E}_{\mathrm{ads}} = \alpha_m
+ \beta_{\mathrm{O}}\,\widetilde{\mathrm{GCN}}_{\mathrm{O}}
+ \beta_{\mathrm{H}}\,\widetilde{\mathrm{GCN}}_{\mathrm{H}}
+ \gamma\,\widetilde{\mathrm{GCN}}_{\mathrm{O}}\,\widetilde{\mathrm{GCN}}_{\mathrm{H}}
+ \varepsilon,
\label{eq:multivar}
\end{equation}
where $\alpha_m$ is a motif-specific intercept ($m \in \{\mathrm{BH, HH, TH}\}$), $\widetilde{\mathrm{GCN}}$ denotes the particle-centered coordination numbers, and the coefficients $\beta_{\mathrm{O}}$, $\beta_{\mathrm{H}}$, and $\gamma$ are shared across motifs.

Second, the unique predictive contribution of each predictor block was evaluated through a drop-one-predictor analysis of 5,933 retained configurations from Ni$_{91}$ (2,654) and Ni$_{130}$ (3,279). Five predictor blocks were considered: adsorption motif, $\mathrm{GCN_O}$, $\mathrm{GCN_H}$, O-site--H-site distance $d_{\mathrm{site\text{-}site}}$, and particle identity. The continuous descriptors were discretized into bins (unit-wide intervals for the GCNs), while the motif and particle identity were treated categorically. The resulting categorical predictor blocks were one-hot encoded and fitted by ordinary least-squares regression. For each predictor block $t$, the unique coefficient of determination was calculated as
\begin{equation}
\Delta \mathrm{R}^2_t = \mathrm{R}^2_{\mathrm{full}} - \mathrm{R}^2_{\mathrm{full}\setminus t},
\label{eq:unique_r2}
\end{equation}
where $\mathrm{R}^2_{\mathrm{full}}$ and $\mathrm{R}^2_{\mathrm{full}\setminus t}$ are the out-of-sample coefficients of determination obtained with all predictor blocks and with block $t$ excluded, respectively. Results were averaged over 15 random $75/25$ train/test splits.

\subsection{Local-environment coverage across systems}
\label{sec:si_coverage}

\begin{figure}[H]
    \centering
    \includegraphics[width=1.0\linewidth]{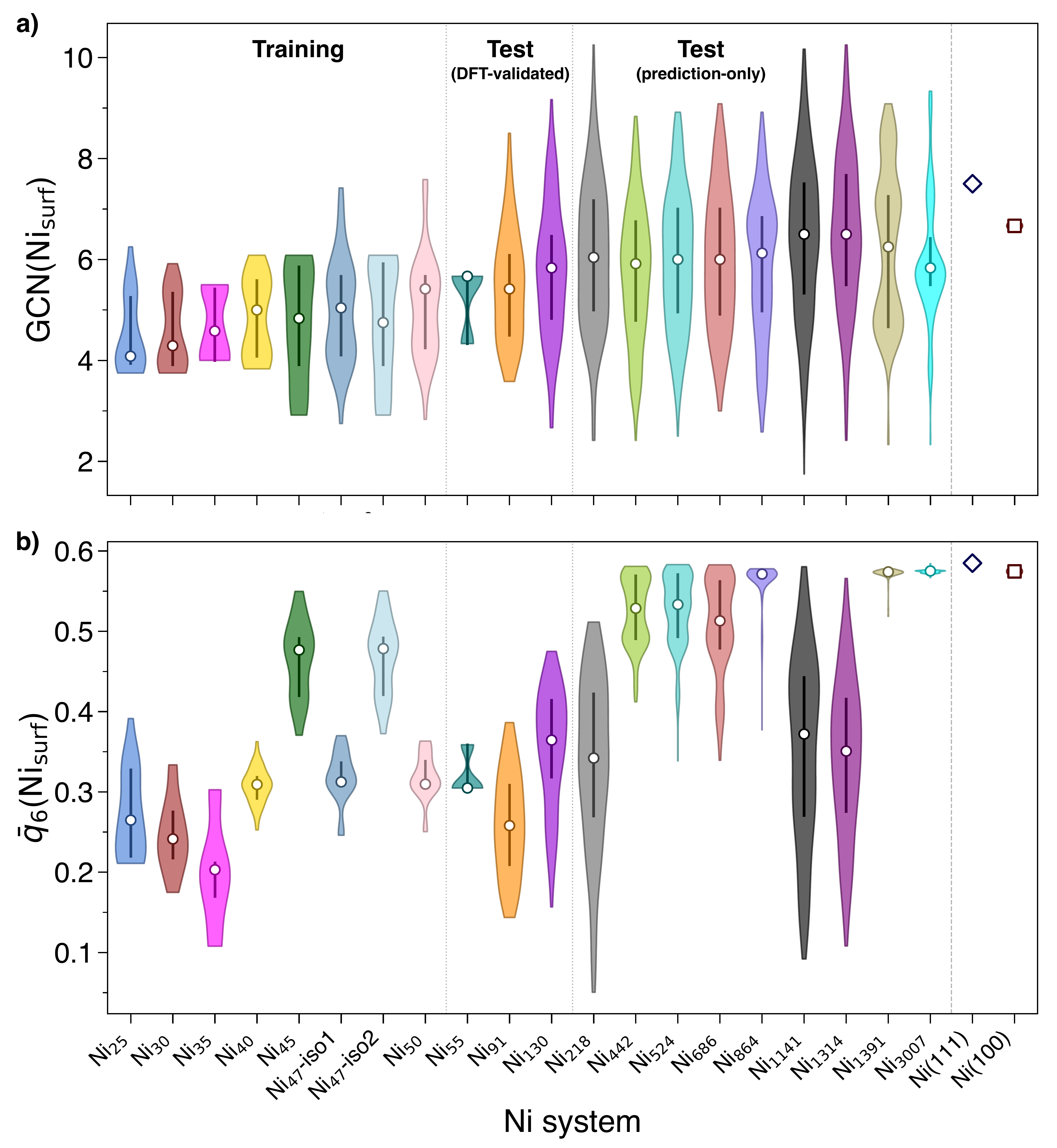}
    \caption{Distributions of the surface (a) atomic generalized coordination number $\mathrm{GCN}$ and (b) averaged bond-orientational order parameter $\bar{q}_6$ of the Ni surface atoms, for all systems considered: the training nanoclusters (Ni$_{25}$--Ni$_{50}$ and the two Ni$_{47}$ isomers), the DFT-validated test nanoclusters (Ni$_{55}$, Ni$_{91}$, Ni$_{130}$), and the prediction-only nanoparticles (Ni$_{218}$--Ni$_{3007}$), together with the Ni(111) and Ni(100) terminations (right).}
    \label{fig:si_coverage}
\end{figure}

\begin{figure}[H]
    \centering
    \includegraphics[width=1.0\linewidth]{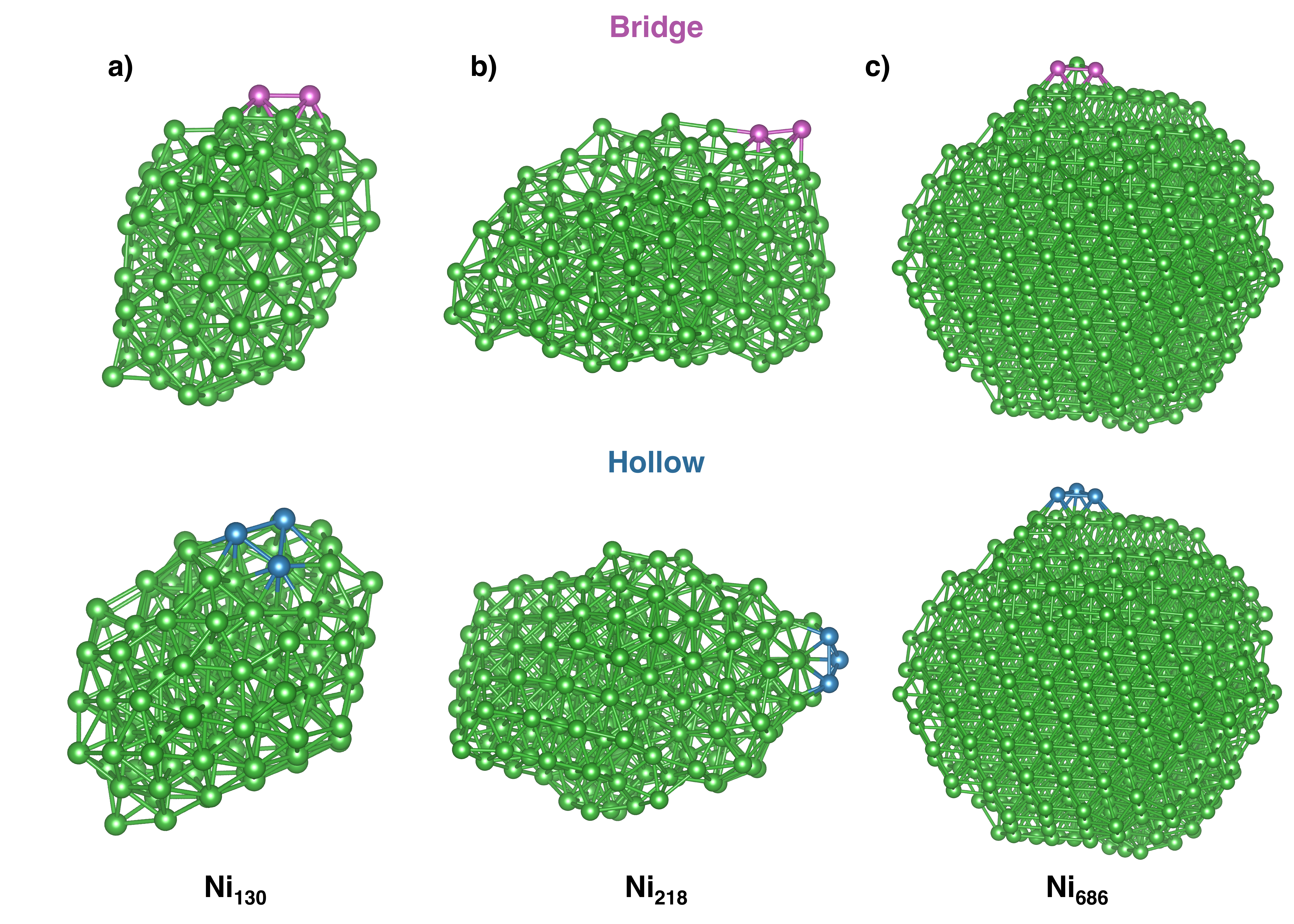}
    \caption{Examples of the lowest predicted-energy bridge (B, top) and threefold-hollow (H, bottom) O-binding sites passing the retention filter for BH and HH co-adsorption, respectively, on (a) Ni$_{130}$, (b) Ni$_{218}$, and (c) Ni$_{686}$. The Ni atoms defining the B and H adsorption-site anchors are highlighted in pink and blue, respectively, while the remaining Ni atoms are shown in green.}
    \label{fig:si_examples_BH_HH}
\end{figure}

\clearpage
\restoregeometry
\doublespacing

\renewcommand{\thepage}{\arabic{page}}
\renewcommand{\theHpage}{REF.\arabic{page}}

\renewcommand{\addcontentsline}[3]{}

\makeatletter
\setcounter{page}{\numexpr\Hy@abspage+1\relax}
\makeatother
\bibliography{bib}

\end{document}